\documentstyle{mn}
\input{epsf.tex}
\newif\ifAMStwofonts

\ifoldfss
 \ifCUPmtlplainloaded \else
  \NewTextAlphabet{textbfit} {cmbxti10} {}
  \NewTextAlphabet{textbfss} {cmssbx10} {}
  \NewMathAlphabet{mathbfit} {cmbxti10} {} 
  \NewMathAlphabet{mathbfss} {cmssbx10} {} 
 \fi %
 \ifAMStwofonts
  \ifCUPmtlplainloaded \else
   \NewSymbolFont{upmath} {eurm10}
   \NewSymbolFont{AMSa} {msam10}
   \NewMathSymbol{\upi}   {0}{upmath}{19}
   \NewMathSymbol{\umu}   {0}{upmath}{16}
   \NewMathSymbol{\upartial}{0}{upmath}{40}
   \NewMathSymbol{\leqslant}{3}{AMSa}{36}
   \NewMathSymbol{\geqslant}{3}{AMSa}{3E}

    \let\le=\leqslant
    \let\ge=\geqslant
  \fi
 \fi
\fi 

\ifnfssone
 \newmathalphabet{\mathit}
 \addtoversion{normal}{\mathit}{cmr}{m}{it}
 \addtoversion{bold}{\mathit}{cmr}{bx}{it}
 \newmathalphabet{\mathbfit} 
 \addtoversion{normal}{\mathbfit}{cmr}{bx}{it}
 \addtoversion{bold}{\mathbfit}{cmr}{bx}{it}
 \newmathalphabet{\mathbfss} 
 \addtoversion{normal}{\mathbfss}{cmss}{bx}{n}
 \addtoversion{bold}{\mathbfss}{cmss}{bx}{n}
 \ifAMStwofonts
  \ifCUPmtlplainloaded \else
   \UseAMStwoboldmath
   \makeatletter
   \new@mathgroup\upmath@group
   \define@mathgroup\mv@normal\upmath@group{eur}{m}{n}
   \define@mathgroup\mv@bold\upmath@group{eur}{b}{n}
   \edef\UPM{\hexnumber\upmath@group}
   \new@mathgroup\amsa@group
   \define@mathgroup\mv@normal\amsa@group{msa}{m}{n}
   \define@mathgroup\mv@bold\amsa@group{msa}{m}{n}
   \edef\AMSa{\hexnumber\amsa@group}
   \makeatother
   \mathchardef\upi="0\UPM19
   \mathchardef\umu="0\UPM16
   \mathchardef\upartial="0\UPM40
   \mathchardef\leqslant="3\AMSa36
   \mathchardef\geqslant="3\AMSa3E

    \let\le=\leqslant
    \let\ge=\geqslant
  \fi
 \fi
\fi 

\ifnfsstwo
 \DeclareMathAlphabet{\mathbfit}{OT1}{cmr}{bx}{it}
 \SetMathAlphabet\mathbfit{bold}{OT1}{cmr}{bx}{it}
 \DeclareMathAlphabet{\mathbfss}{OT1}{cmss}{bx}{n}
 \SetMathAlphabet\mathbfss{bold}{OT1}{cmss}{bx}{n}
 \ifAMStwofonts
  \ifCUPmtlplainloaded \else
   \DeclareSymbolFont{UPM}{U}{eur}{m}{n}
   \SetSymbolFont{UPM}{bold}{U}{eur}{b}{n}
   \DeclareSymbolFont{AMSa}{U}{msa}{m}{n}
   \DeclareMathSymbol{\upi}{0}{UPM}{"19}
   \DeclareMathSymbol{\umu}{0}{UPM}{"16}
   \DeclareMathSymbol{\upartial}{0}{UPM}{"40}
   \DeclareMathSymbol{\leqslant}{3}{AMSa}{"36}
   \DeclareMathSymbol{\geqslant}{3}{AMSa}{"3E}

    \let\le=\leqslant
    \let\ge=\geqslant
  \fi
 \fi
\fi 

\ifCUPmtlplainloaded \else
 \ifAMStwofonts \else 
  \def\upi{\pi}
  \def\umu{\mu}
  \def\upartial{\partial}
 \fi
\fi

\title{Chaos and chaotic phase mixing in cuspy triaxial potentials}

\author[H. E. Kandrup and C. Siopis]
 {Henry E. Kandrup$^{1,2,3}$\thanks{E-mail: kandrup@astro.ufl.edu}
  and Christos Siopis$^{4}$\thanks{E-mail: siopis@umich.edu}\\
  $^{1}$ Department of Astronomy, University of Florida, Gainesville, 
     FL 32611-2055, USA\\
  $^{2}$ Department of Physics, University of Florida, Gainesville, 
     FL 32611-2055, USA\\
  $^{3}$ Institute for Fundamental Theory, University of Florida, Gainesville,
     FL 32611-2055, USA\\
  $^{4}$ Department of Astronomy, University of Michigan, Ann Arbor, MI
	 48109-1090, USA}

\date{Accepted 2003 \hskip 1in .
   Received 2002 \hskip 1in .}

\pagerange{\pageref{firstpage}--\pageref{lastpage}}
\pubyear{2002}

\begin{document}

\maketitle

\label{firstpage}

\begin{abstract}

This paper continues an investigation of chaos and chaotic phase mixing in triaxial generalisations of the Dehnen potential which have been proposed to describe realistic ellipticals that have a strong density cusp and manifest significant deviations from axisymmetry. Earlier work is extended in three important ways, namely by exploring systematically the effects of (1) variable axis ratios, (2) `graininess' associated, e.g., with stars and bound substructures, idealised as friction and white noise, and (3) large-scale organised motions within a galaxy and a dense cluster environment, each presumed to induce near-random forces idealised as coloured noise with a finite autocorrelation time. The effects of varying the axis ratio were studied in detail by considering two sequences of models with cusp exponent ${\gamma}=1$ and, respectively, axis ratios $a : b : c = 1.00 : 1.00 - {\Delta} : 0.50$ and $a : b : c = 1.00 : 1.00 - {\Delta} : 1.00 - 2{\Delta}$ for variable ${\Delta}$. Three important conclusions are (1) that not all the chaos can be attributed to the presence of the cusp, (2) that significant chaos can persist even for axisymmetric systems, and (3) that the introduction of a supermassive black hole can induce both moderate increases in the relative number of chaotic orbits and substantial increases in the size of the largest Lyapunov exponent. In the absence of any perturbations, the coarse-grained distribution function associated with an initially localised ensemble of chaotic orbits evolves exponentially towards a nearly time-independent form at a rate ${\Lambda}$ that correlates with the typical values of the finite time Lyapunov exponents ${\chi}$ associated with the evolving orbits. Allowing for discreteness effects and/or an external environment accelerates phase space transport {\it both} by increasing  the rate at which orbits spread out within a given phase space region {\it and} by facilitating diffusion along the Arnold web that connects different phase space regions, so as to facilitate an approach towards a true equilibrium. The details of the perturbation appear unimportant. All that really matters are the amplitude and, for the case of coloured noise, the autocorrelation time, {\em i.e.,} the characteristic time over which the perturbation varies. Overall the effects of the perturbations scale logarithmically in both amplitude and autocorrelation time. Even comparatively weak perturbations can increase ${\Lambda}$ by a factor of three or more, a fact that has potentially significant implications for violent relaxation.

\end{abstract}


\section{MOTIVATION}

The first paper in this series (Siopis \& Kandrup 2000), hereafter denoted Paper I, began an investigation of phase space transport and chaotic phase mixing in triaxial generalisations of the Dehnen (1993) potentials which have been proposed (e.g., Merritt \& Fridman 1996) to model realistic elliptical galaxies that have a strong density cusp and manifest significant deviations from axisymmetry. These correspond to potentials associated self-consistently with the mass density

\begin{equation}
{\rho}(m)={(3-{\gamma})\over 4{\pi}abc}m^{-{\gamma}}(1+m)^{-(4-{\gamma})},
\end{equation}
where
\begin{equation}
m^{2}={x^{2}\over a^{2}}+{y^{2}\over b^{2}}+{z^{2}\over c^{2}}.
\end{equation}

As discussed in Paper I, this potential was chosen primarily to facilitate comparisons with, and to improve the relevance of, the results derived here with those of other workers who have been using it. It is thought to constitute a quasi-realistic approximation for the central regions of many elliptical galaxies, at least compared with St\"ackel-type potentials which lack a central density cusp. Although there is no guarantee that there exist exact self-consistent equilibria corresponding to triaxial Dehnen mass distributions, insights gained from a study of motion in this potential can provide clues towards an understanding of more generic cuspy triaxial potentials and, importantly, of their evolution towards a (quasi-)equilibrium state. 

Paper I explored the effects of varying ${\gamma}$, which controls the steepness of the density cusp, and introducing a central supermassive black hole of variable mass, as well as allowing for low amplitude perturbations intended to mimic the effects of discreteness effects (i.e., gravitational Rutherford scattering) and/or an external environment. However, that paper was incomplete in that it did not allow for the effects of variable axis ratios, attention focusing exclusively on the values $c/a=1/2$ and $(a^{2}-b^{2})/(a^{2}-c^{2})=1/2$ considered originally by Merritt \& Fridman (1996). 

That work was also incomplete in that the modeling of external perturbations was very simplistic. Most of that modeling focused on the effects of periodic driving, in which the potential is subjected to a time-dependent periodic perturbation characterised by at most three different frequencies. Although this might seem reasonable when considering the effects of a single large companion or a few smaller satellite galaxies, this is certainly not appropriate to model a rich cluster environment, where galaxies tend to be much closer. When considering a dense environment one must allow for more complex perturbations which, presumably, are far from periodic, perturbations which, as described in Paper I, would seem better modeled as random kicks of finite duration, i.e., coloured noise. Paper I did indeed describe a small number of experiments involving coloured noise, but these were far from exhaustive. In particular, no effort was made to determine the extent to which the detailed form of the perturbation actually matters: all the simulations involved perturbations idealised as an Ornstein-Uhlenbeck process (cf. van Kampen 1981).

The work on chaotic phase mixing described in Paper I was also incomplete in the sense that the discussion was largely qualitative. It was observed that, as for other potentials (cf. Kandrup \& Mahon 1994, Kandrup 1998b), ensembles of chaotic orbits typically exhibit a rapid evolution towards a near-equilibrium (a near-invariant distribution), but it was not confirmed explicitly that this evolution proceeds exponentially in time. Moreover, there was no systematic exploration as to how, for unperturbed systems, the rate ${\Lambda}$ at which this evolution proceeds correlates with the size of a typical Lyapunov exponent ${\chi}$; or, for perturbed systems, how ${\Lambda}$ correlates with the amplitude of the perturbation.

The aim of this second paper is to fill these remaining lacunae. Section 2 summarises an investigation of how, neglecting discreteness and environmental effects but allowing for a central supermassive black hole, changes in the axis ratios affect (i) the relative number of regular and chaotic orbits, (ii) the typical sizes of the Lyapunov exponents, and (iii) the overall efficacy of phase space transport. It was found that, in general, larger deviations from axisymmetry or spherical symmetry tend to increase the fraction of chaotic orbits, although considerable chaos can arise even in axisymmetric systems, especially at higher energies. Alternatively, strongly triaxial systems do not in general tend to have Lyapunov exponents that are much larger than moderately triaxial systems. The introduction of a black hole increases significantly the size of the largest Lyapunov exponent, both in absolute units and units of the dynamical time $t_{D}$, but does not in general result in a very much larger measure of chaotic orbits. The work described here does not address the important issue of the relative abundances of different types of regular orbits. However, this issue is currently under investigation in a more general setting (Kandrup \& Siopis 2003).

Section 3 focuses on chaotic phase mixing. An analysis of coarse-grained distribution functions is used to confirm that ensembles of orbits typically evolve exponentially towards a near-invariant distribution and that, in the absence of all perturbations, the rate at which this evolution proceeds correlates with the typical size of the finite time Lyapunov exponents associated with the ensemble. Discreteness effects, modeled (cf. Chandrasekhar 1943) as friction and white noise, can dramatically accelerate chaotic phase mixing {\it both} by increasing the rate at which orbits spread out within a given chaotic phase space region {\it and} by facilitating diffusion along the Arnold web that connects different phase space regions. 

Section 3 also focuses on how chaotic phase mixing is impacted by large-scale organised motions within a galaxy or a dense cluster environment, each modeled as coloured noise. One principal conclusion of the analysis, consistent with Paper I, is that the details of the perturbations seem to be largely immaterial: The only things that really seem to matter are (1) the amplitude and (2) the autocorrelation time (i.e., the characteristic time scale associated with individual kicks), both of which can be readily estimated via dimensional analysis. In other words, {\it the effects of the environment seem to be insensitive to details which are difficult to ascertain observationally!} The other principal conclusion is that choices of amplitude and autocorrelation time appropriate for real galaxies can in fact lead to significant effects on time scales short compared with the age of the Universe.

Section 4 summarises the principal conclusions and Section 5 discusses potential implications for real galaxies.

\section[]{CHAOS AS A FUNCTION OF AXIS RATIO\\* AND BLACK HOLE MASS}

\subsection{The numerical experiments}

In order to understand the effects of variable axis ratio, two different sequences of models were considered in detail. The first involved sweeping through a variety of triaxial configurations which connected prolate and oblate spheroids. Specifically, this entailed assuming axis ratios $a:b:c$ = $1.00:1.00-{\Delta}:0.50$, with ${\Delta}$ allowed to vary between $0.00$ and $0.50$ by increments ${\delta}=0.05$, but also considering several extra models with ${\Delta}$ closer to $0.00$ and $0.50$. The second sequence involved triaxial deviations from a spherical system, corresponding to axis ratios $a:b:c$ = $1.00:1.00-{\Delta}:1.00-2{\Delta}$. 

Attention was restricted primarily to models with cusp index ${\gamma}=1$ but three different black hole masses were considered, namely $M_{BH}=0$, $M_{BH}=10^{-3}$, and $M_{BH}=10^{-2}$.

The principal focus was on determining the statistical properties of orbit ensembles as a function of energy $E$. This was done by considering ten different energies, $-1.0{\;}{\le}{\;}E{\;}{\le}{\;}-0.1$, and, for each energy, selecting 1000 `representative' initial conditions. (The model with axis ratio $1.00:0.75:0.50$, which arises in both sequences, was studied for two different sets of $1000$ initial conditions and it was confirmed that the conclusions were in agreement statistically.) Arguably the most honest sampling of any given energy entails a uniform sampling of the constant energy hypersurface; and, for this reason, orbits were selected to sample the microcanonical distribution
\begin{equation}
{\mu}{\;}{\propto}{\;}{\delta}_{D}(H-E),
\end{equation}
with $H$ the Hamiltonian. This distribution turns out to be difficult and expensive to sample directly, especially since the potential $V$ cannot be expressed analytically. For this reason an indirect approach was used. By integrating over the dependence on velocity, it is easily seen that a microcanonical distribution corresponds to a configuration space distribution
\begin{equation}
f({\bf r}){\;}{\propto}{\;}\cases{ \left( E-V \right)^{1/2} & if
$V({\bf r}){\;}{\le}{\;}E$;\cr 
 & \cr
0 & if $V({\bf r})>E$.\cr} 
\end{equation}
Alternatively, the velocity distribution at any given configuration space is isotropic, with each mass having a speed $v=\sqrt{2(E-V)}$. To obtain a random, uniform sampling of the constant energy surface, it therefore suffices to (1) sample $f({\bf r})$ to generate a collection of 1000 configuration space points and (2) to each point assign a velocity of magnitude $v=\sqrt{2(E-V)}$ oriented in a randomly chosen direction.

Each orbit was integrated for a time ${\ge}{\;}200t_{D}$ using a variable timestep integrator with accuracy parameter $10^{-8}$ which, in every case, conserved energy to at least $1$ part in $10^{5}$. Estimates of the largest (finite time) Lyapunov exponent were obtained by tracking the evolution of a nearby orbit which was periodically renormalised in the usual way. The dynamical time $t_{D}$ for a given choice of energy and axis ratios was identified as follows: For the orbits in each ensemble, define $t_{x}$ -- and, analogously, the quantities $t_{y}$ and $t_{z}$ -- as the mean time between successive crossings of the $x=0$ plane; and, given $t_{x}$, $t_{y}$, and $t_{z}$, define 
\begin{equation}
t_{D}=2{\pi}(t_{x}+t_{y}+t_{z}).
\end{equation}
This definition seems well motivated physically; the choice of normalisation factor $2{\pi}$ ensured that, for the `maximally triaxial' model first considered in detail by Merritt and Fridman (1996), this definition of $t_{D}$ agreed with the Merritt-Fridman definition for all energies to within $3\%$.

The degree of chaos manifested by the different ensembles was quantified using two complementary diagnostics, namely finite time Lyapunov exponents (Grassberger et al 1988), which probe the degree of exponential sensitivity exhibited by different orbits, and orbital complexities (Kandrup, Eckstein, and Bradley 1997), which probe the extent to which the power associated with an individual orbit is concentrated at or near a few special frequencies. This entailed determining for each orbit the quantities $n_{x}$, $n_{y}$, and $n_{z}$, defined, respectively, as the minimum number of frequencies required to capture a fixed fraction $k$ of the power in each direction, and then assigning a total complexity
\begin{equation}
n=n_{x}+n_{y}+n_{z}.
\end{equation}
Given temporal discreteness effects reflecting the fact that the orbital data were sampled at intervals ${\approx}{\;}0.25t_{D}$, the cleanest distinctions between different orbits were obtained for $k{\;}{\approx}{\;}0.9$. 

The degree to which individual chaotic orbits are `sticky' (Contopoulos 1971), i.e., that they can remain `stuck' in a small part of the accessible phase space for a comparatively long time, was probed by computing finite time Lyapunov exponents for long time integrations of several different initial conditions on the same phase space hypersurface and determining the time scale (or, in some cases, a lower bound) on which the exponents for different orbits converge towards a single asymptotic value.

Because it can take an orbit a very long time to uniformly sample the accessible phase space, it can be difficult and expensive computationally to compute estimates of the largest Lyapunov exponent as a function of axis ratio, black hole mass, and energy. For this reason, such estimates were instead obtained typically by computing the mean value of the finite time Lyapunov exponents for the chaotic orbits in a given $1000$ orbit ensemble, a procedure which is justified theoretically given the assumption of ergodicity. For the case of models where only a very few orbits are chaotic, this procedure is suspect statistically, so that the value of the largest Lyapunov was computed by averaging over the results of extremely long time integrations of four chaotic initial conditions.

Uncertainties in the relative measure of chaotic orbits were estimated assuming $N^{1/2}$ statistics. Uncertainties in the estimated Lyapunov exponent were obtained by analysing separately the two halves of the 1000 orbit ensembles.

\subsection{Results}

\begin{figure}
\centering
\centerline{
    \epsfxsize=8cm
    \epsffile{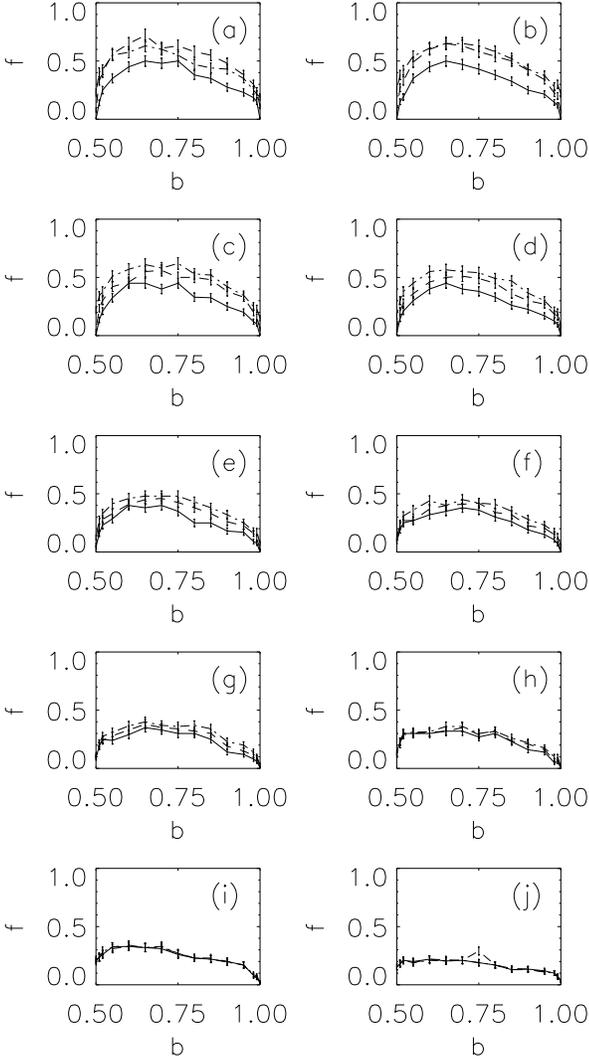}
      }
    \begin{minipage}{10cm}
    \end{minipage}
    \vskip -0.3in\hskip -0.0in
\caption{
(a) The fraction of orbits with energy $E=-1.0$ that are chaotic for models
with axis ratios $1.00:b:0.50$ and black hole mass
$M_{BH}=0$ (solid curve), $M_{BH}=10^{-3}$ (dashed curve), and 
$M_{BH}=10^{-2}$ (dot-dashed curve)
(b) The same for $E=-0.9$.
(c) The same for $E=-0.8$.
(d) The same for $E=-0.7$.
(e) The same for $E=-0.6$.
(f) The same for $E=-0.5$.
(g) The same for $E=-0.4$.
(h) The same for $E=-0.3$.
(i) The same for $E=-0.2$.
(j) The same for $E=-0.1$.
}
\label{landfig}
\end{figure}

Consider first the sequence of models extending between oblate and prolate axisymmetric configurations. Here the most obvious point is that, overall, triaxial models which manifest larger deviations from axisymmetry tend to admit larger measures of chaotic orbits. This is, e.g. evident from FIGURE 1, which exhibits the fraction $f$ of chaotic orbits as a function of the intermediate axis $b$
for ten different energies.

Nevertheless, as is evident from FIGURE 1, even axisymmetric configurations can admit significant measures of chaos, at least at higher energies. This chaos appears to reflect the large scale structure of the bulk potential, not the presence of a central cusp. Indeed, for very high energies, one finds virtually identical numbers for a cuspless ${\gamma}=0$ model although, for lower energies, the ${\gamma}=0$ admits a much smaller measure of chaotic orbits. This is illustrated in FIGURE 2 which, for a model with axis ratio $1.00:0.50:0.50$, exhibits the ordered values of the 1000 computed Lyapunov exponents for both ${\gamma}=1$ and ${\gamma}=0$ for energies ranging between $E=-0.1$ and $E=-0.5$. However, despite this chaos at higher energies, except for the case of prolate axisymmetric configurations there tends overall to be more chaos at lower energies, presumably associated with the cusp: this chaos is, e.g., absent for analogous models with ${\gamma}=0$. For a broad range of parameter values, the measure of chaotic orbits ranges between about $10\%$ and $50\%$.
\begin{figure}
\centering
\centerline{
    \epsfxsize=8cm
    \epsffile{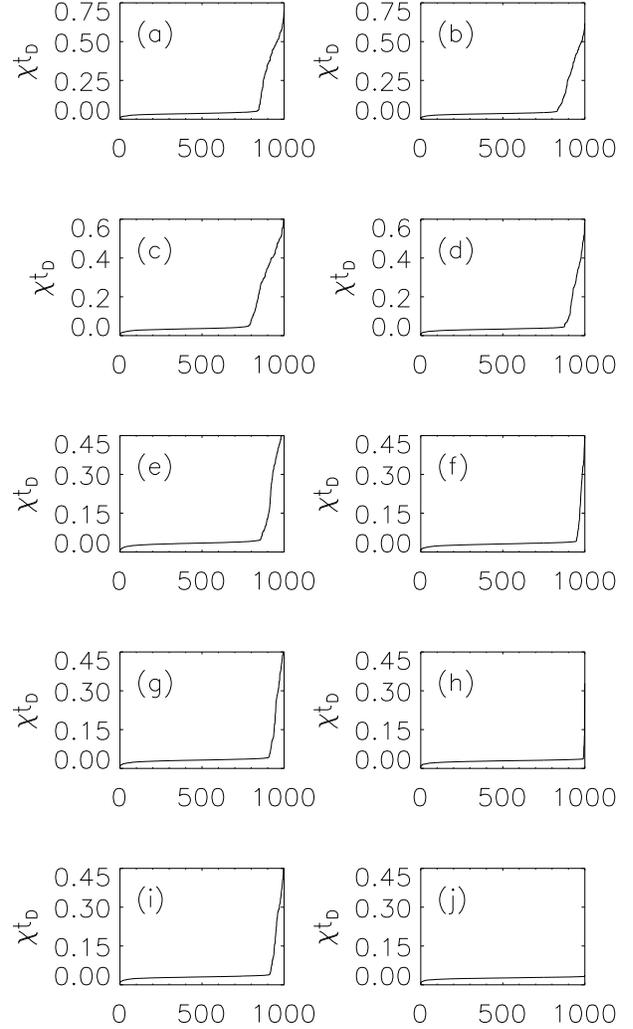}
      }
    \begin{minipage}{10cm}
    \end{minipage}
    \vskip -0.3in\hskip -0.0in
\caption{
(a) An ordered plot of the computed Lyapunov exponents for the ${\gamma}=1$
model with $a:b:c=1.00:0.50:0.50$ and energy $E=-0.1$. (b) The same for 
${\gamma}=0$.
(c) and (d) The same for $E=-0.2$.
(e) and (f) The same for $E=-0.3$.
(g) and (h) The same for $E=-0.4$.
(i) and (j) The same for $E=-0.5$.
}
\label{landfig}
\end{figure}

That chaos was observed in the prolate model with axis ratio $1.00:0.50:0.50$ but not in the oblate model with $1.00:1.00:0.50$ raises the question as to whether chaos is in fact generic in the axisymmetric Dehnen potentials. This issue was addressed by considering a variety of prolate and oblate models with, respectively, axis ratios $1.00:1.00-{\Delta}:1.00-{\Delta}$ and $1.00:1.00:1.00-{\Delta}$. The net result is that chaos can arise for both prolate and oblate systems, but that, for the case of oblate systems, a substantially larger deviation from sphericality is required. For the sequence with $1.00:1.00-{\Delta}:1.00-{\Delta}$, a significant amount of chaos, ${\sim}{\;}2\%$, is observed already at energy $E=-0.1$ for ${\Delta}=0.2$; for the sequence with $1.00:1.00:1.00-{\Delta}$, one requires a value as large as ${\Delta}=0.6$ to find a comparable amount of chaos. This is illustrated in FIGURE 3, the top two panels of which exhibit the fraction $f$ of chaotic orbits with energy $E=-0.1$ for different choices of axis ratio. The lower two panels exhibit estimates of the largest Lyapunov exponent for the same configurations. In each case, the onset of chaos is characterised by very small Lyapunov exponents which, however, increases monotonically or near-monotonically with increasing ${\Delta}$.
\begin{figure}
\centering
\centerline{
    \epsfxsize=8cm
    \epsffile{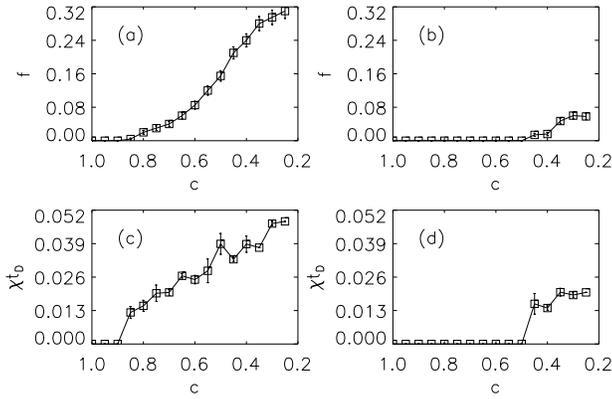}
      }
    \begin{minipage}{10cm}
    \end{minipage}
    \vskip -1.3in\hskip -0.0in
\caption{
(a) The fraction $f$ of chaotic orbits for $E=-0.1$ and variable axis
ratios $a:b:c = 1.00\,:\,c\,:\,c$.
(b) The same for $a:b:c=1.00\,:\,1.00\,:\,c$.
(c) The largest Lyapunov exponent for the parameter values in (a).
(d) The same (b).
}
\label{landfig}
\end{figure}

It is interesting that an analogous result has been obtained for the so-called thermal equilibrium model, a standard pedagogical example (cf. Brown \& Reiser 1995) from the physics of charged particle beams. For this system, which corresponds to a self-interacting nonneutral plasma in thermal equilibrium confined by an anisotropic harmonic oscillator potential, one finds (Bohn \& Sideris 2003) that even slightly prolate configurations tend to admit large amounts of chaos, whereas comparably aspherical oblate configurations exhibit little, if any, chaos.

The remaining interesting point is that, even when an axisymmetric model admits no chaotic orbits, relatively small perturbations away from axisymmetry suffice to trigger a significant amount of chaos, $10\%$ or more. This is in qualitative agreement with recent computations by El-Zant and Shlosman (2002), who explored the effects of bars with variable amplitude on orbits in an otherwise axisymmetric potential. 
\begin{figure}
\centering
\centerline{
    \epsfxsize=8cm
    \epsffile{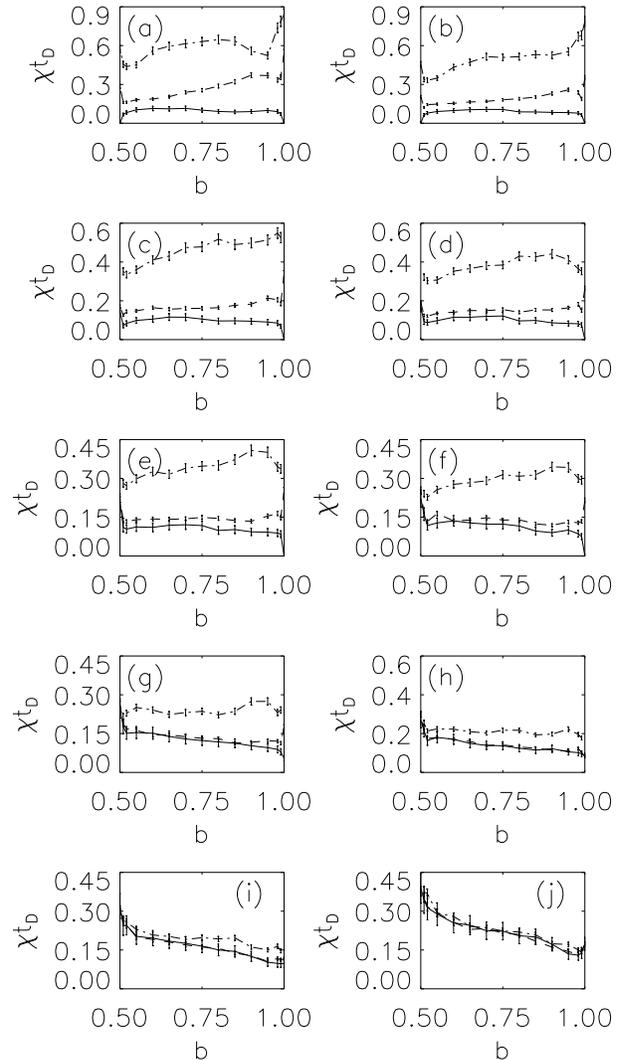}
      }
    \begin{minipage}{10cm}
    \end{minipage}
    \vskip -0.3in\hskip -0.0in
\caption{
(a) Estimates of the largest Lyapunov exponent for chaotic orbits with energy 
$E=-1.0$ for models with axis ratio $1.00:b:0.50$ and black hole
mass $M_{BH}=0$ (solid curve), $M_{BH}=10^{-3}$ (dashed curve), and 
$M_{BH}=10^{-2}$ (dot-dashed curve)
(b) The same for $E=-0.9$.
(c) The same for $E=-0.8$.
(d) The same for $E=-0.7$.
(e) The same for $E=-0.6$.
(f) The same for $E=-0.5$.
(g) The same for $E=-0.4$.
(h) The same for $E=-0.3$.
(i) The same for $E=-0.2$.
(j) The same for $E=-0.1$.
}
\label{landfig}
\end{figure}

Overall, when scaled in physical units, the size of the largest Lyapunov exponent ${\chi}t_{D}{\;}{\sim}{\;}0.2$, {\em i.e.,} the growth time is roughly $5t_{D}$. This is, e.g. evident from FIGURE 4 which exhibits estimates of the largest Lyapunov exponent for the same ensembles used to generate FIGURE 1. For lower energies, this conclusion is again comparatively insensitive to the choice of axis ratio. However, the behaviour at much higher energies, where chaos is triggered by the bulk potential, is different. In this case, the value of ${\chi}t_{D}$ is actually {\em maximised} for the axisymmetric model with $1.00:0.50:0.50$ and decreases monotonically as one moves along the sequence towards $1.00:1.00:0.50$.

\begin{figure}
\centering
\centerline{
    \epsfxsize=8cm
    \epsffile{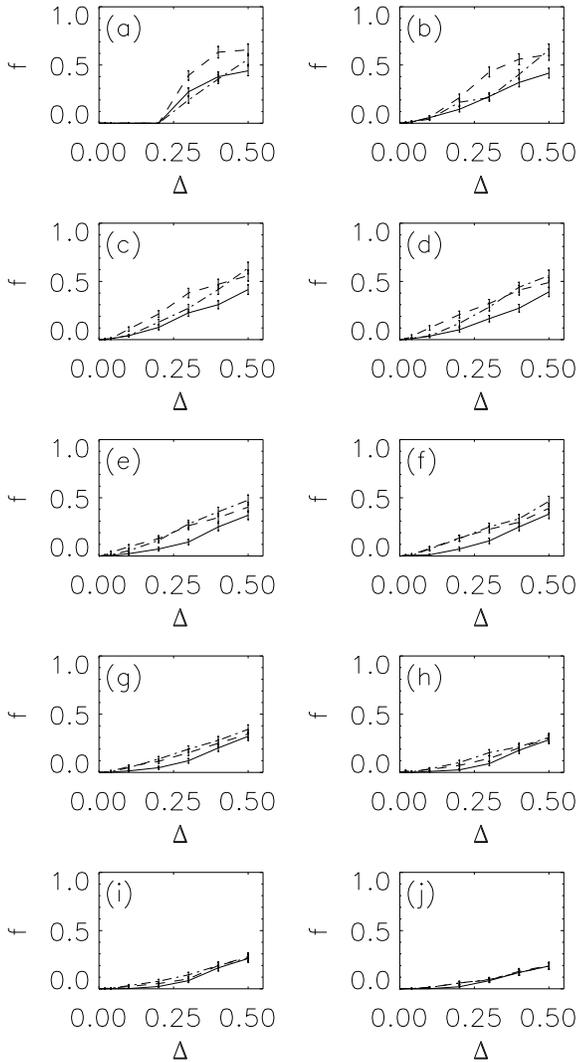}
      }
    \begin{minipage}{10cm}
    \end{minipage}
    \vskip -0.3in\hskip -0.0in
\caption{
(a) The fraction of orbits with energy $E=-1.0$ that are chaotic for models
with axis ratios $1.00:b:c$ for $b=1-{\Delta}$ and $c=1-2{\Delta}$, and black 
hole mass
$M_{BH}=0$ (solid curve), $M_{BH}=10^{-3}$ (dashed curve), and 
$M_{BH}=10^{-2}$ (dot-dashed curve)
(b) The same for $E=-0.9$.
(c) The same for $E=-0.8$.
(d) The same for $E=-0.7$.
(e) The same for $E=-0.6$.
(f) The same for $E=-0.5$.
(g) The same for $E=-0.4$.
(h) The same for $E=-0.3$.
(i) The same for $E=-0.2$.
(j) The same for $E=-0.1$. The unusual appearance of panel (a) reflects the
fact that, for nearly spherical models, the minimum energy is somewhat
larger than $E=-1.0$.
}
\label{landfig}
\end{figure}

The introduction of a black hole has only a minimal effect on high energy orbits which spend little time in the central region where they can `feel' the gravitational influence of the hole. At lower energies, however, the presence of a black hole tends to increase the relative measure of chaotic orbits, albeit not by all that much. Thus, e.g. as is evident from FIGURE 1, the introduction of a hole with $M_{BH}=10^{-2}$ invariably increases the fraction $f$ by $50\%$ or less. There always remains a significant measure of regular orbits. Interesting also is the fact that curves of $f({\Delta})$ for models with and without a black hole manifest similar curvatures: the only obvious difference is that, for $M_{BH}=0$, there are somewhat fewer chaotic orbits.

Although a black hole changes $f$ by only a relatively small amount, it typically occasions a substantial increase in the value of the largest Lyapunov exponent for low energies. Moreover, one sees that although $f$ appears to vary smoothly with shape, the value of the largest Lyapunov exponent can exhibit a more complex dependence on ${\Delta}$. To summarise: the introduction of a black hole typically does {\em not} result in a huge increase in the number of chaotic orbits, but it {\em does} increase significantly the degree of exponential sensitivity exhibited by those orbits which are chaotic.

\begin{figure}
\centering
\centerline{
    \epsfxsize=8cm
    \epsffile{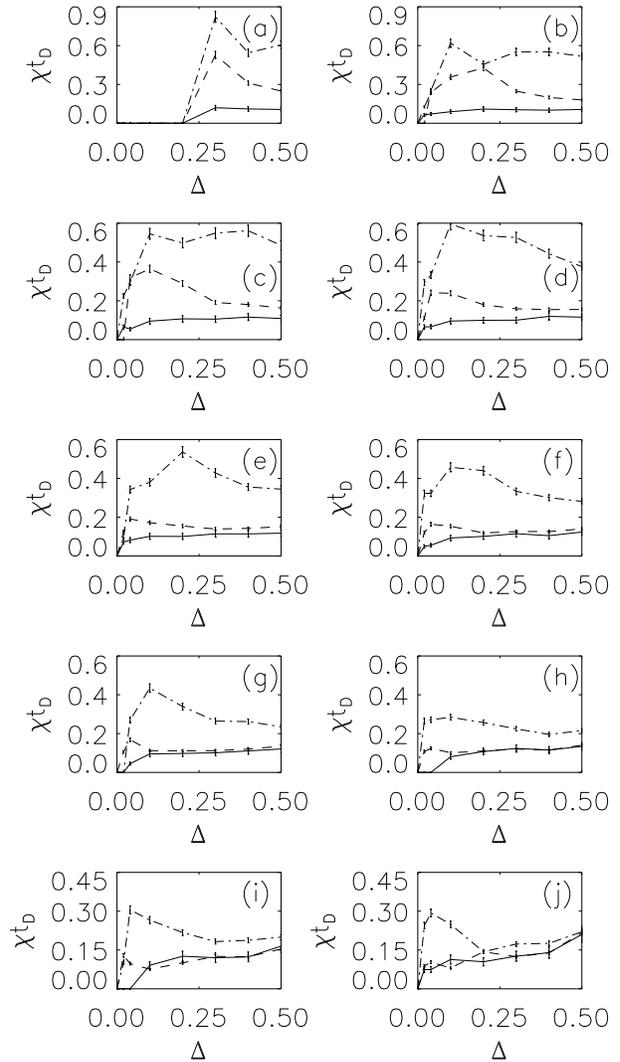}
      }
    \begin{minipage}{10cm}
    \end{minipage}
    \vskip -0.3in\hskip -0.0in
\caption{
(a) Estimates of the largest Lyapunov exponent for chaotic orbits with energy 
$E=-1.0$ for models with axis ratio $1.00:b:c$ for $b=1-{\Delta}$,
and $c=1-2{\Delta}$, and black hole
mass $M_{BH}=0$ (solid curve), $M_{BH}=10^{-3}$ (dashed curve), and 
$M_{BH}=10^{-2}$ (dot-dashed curve)
(b) The same for $E=-0.9$.
(c) The same for $E=-0.8$.
(d) The same for $E=-0.7$.
(e) The same for $E=-0.6$.
(f) The same for $E=-0.5$.
(g) The same for $E=-0.4$.
(h) The same for $E=-0.3$.
(i) The same for $E=-0.2$.
(j) The same for $E=-0.1$.
}
\label{landfig}
\end{figure}

A consideration of the sequence which starts from spherical and becomes triaxial also yields several interesting conclusions. Most obvious from FIGURE 5 is the fact that, for all energies, the relative measure of chaotic orbits is a monotonically increasing function of ${\Delta}$, {\em i.e.,} the deviation from sphericality. Even a deviation as small as ${\Delta}=0.01$ suffices to trigger a nonzero measure of chaotic orbits. However, a comparatively large ${\Delta}$ is required to trigger as much as (say) 20\% chaotic orbits.

Overall, as is evident from FIGURE 6, at least for $M_{BH}=0$ the value of the largest Lyapunov exponent also appears to be a monotonically increasing function of ${\Delta}$. However, especially for lower energies the increase is relatively minimal for ${\Delta}>0.1$ or so. In other words, larger deviations from sphericality yield a larger measure of chaotic orbits, but the degree of exponential sensitivity, as probed by the largest Lyapunov exponent, does not change all that much.

As for the other sequence, the addition of a black hole again tends to increase the abundance of chaotic orbits, albeit not by all that much. Adding a black hole also leads to larger Lyapunov exponents but, unlike the case when $M_{BH}=0$, this exponent is {\em not} a strictly monotonically increasing function of ${\Delta}$. For the larger black hole mass, $M_{BH}=10^{-2}$, the largest exponent increases rapidly with ${\Delta}$ up to a value ${\Delta}{\;}{\approx}{\;}0.1-0.2$ but then becomes a comparatively flat function of ${\Delta}$.

\begin{figure}
\centering
\centerline{
    \epsfxsize=8cm
    \epsffile{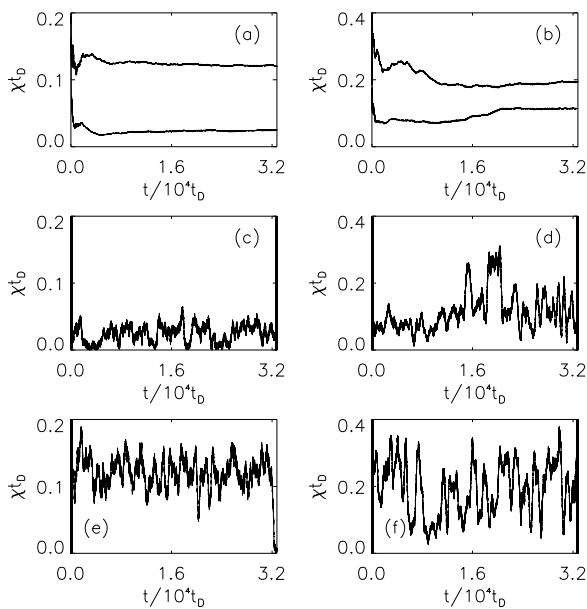}
      }
    \begin{minipage}{10cm}
    \end{minipage}
    \vskip -0.3in\hskip -0.0in
\caption{
(a) Time-dependent estimates of the true (cumultative) Lyapunov exponent for 
two orbits
with $E=-1.0$ evolved in the absence of a black hole in a model with axis 
ratio $1.00:0.95:0.50$.
(b) The same for $E=-0.1$. (c) and (e). Estimates of finite time Lyapunov
exponents for the orbits exhibited in (a), generated by partitioning the
data into a large numer of segments and analysing them individually. (d) 
and (f) The same for the orbits in (b).
}
\label{landfig}
\end{figure}

For virtually all choices of axis ratio, some chaotic orbit segments were so nearly regular that it was highly nontrivial to distinguish then from regular segments. (For this reason distinctions between regular and chaotic were made by using both Lyapunov exponents and complexities as complementary diagnostics.) Moreover, the structure of the chaotic phase space regions is often complex in the sense that two chaotic segments evolved in the same potential with the same energy can have finite time Lyapunov exponents with significantly different values for surprisingly long times. 

This sort of stickiness appears to be especially pronounced for systems that are comparatively close to axisymmetric. One example is exhibited in FIGURE 7, which was generated from orbit segments evolved in a model with axis ratio $1.00:0.95:0.50$. Each of the top two panels shows the time-dependent ${\chi}(t)$, generated in the usual way by also tracking the evolution of a nearby orbit periodically renormalised, for two orbits each with energies $E=-1.0$ and $E=-0.1$ evolved for a total time $t=32768t_{D}$. In each case, estimates of the largest Lyapunov exponent were computed at intervals of $1t_{D}$. The fact that the curves are not decaying uniformly towards zero is {\em prima facia} evidence that the orbits are not regular. How the actual degree of exponential sensitivity varies with time can be gauged from the lower four panels, which exhibit finite time Lyapunov exponents for segments of these orbits. These panels were constructed by extracting from each orbit a collection of $32768$ finite time Lyapunov exponents $\{{\chi}(t_{i})\}$ for intervals $t_{i}<t<t_{i}+1$ ($i=0,32767$) and then smoothing the resulting data by performing a boxcar average over 400 adjacent points. It is evident that, in each case, these finite time exponents exhibit considerable variability. However, it is also clear that the orbits for which the cumultative ${\chi}(t)$ is smaller tend to spend a considerable amount of time in phase space regions where the degree of exponential sensitivity is very small, whereas the orbits with larger ${\chi}$ tend systematically to avoid these regions.

\section{CHAOTIC PHASE MIXING}

\subsection{The numerical experiments}

As in Paper I, the experiments involving chaotic phase mixing entailed (i) selecting localised ensembles of $1600$ initial conditions all with the same energy $E$, (ii) evolving each orbit in the ensemble for a time $t=200t_{D}$, (iii) determining the extent to which the evolving ensemble exhibited a coarse-grained evolution towards a nearly time-independent state, i.e., a near-invariant distribution, and (iv) determining how this evolution was affected by discreteness effects and/or an external environment, modeled as friction and white or coloured noise. 
Attention focused primarily on weak perturbations, characterised typically by a relaxation time $t_{R}$ much longer than the time scales of interest, so that, in a first approximation, the energies of individual orbits are nearly conserved and one can view the perturbations as simply accelerating diffusion on constant energy hypersurfaces (compare Weinberg 2001a,b).

A real galaxy is of course characterised by a complex, many-body potential;
and, for that reason, it is not completely obvious that constructs like 
cantori (Mather 1982) or the Arnold web (Arnold 1964), which can be proven
rigorously for smooth potentials, are applicable to galaxies (see, e.g.,
Contopoulos 2002 for a pedagogical discussion of cantori and the Arnold web). 
However, recent work (Sideris \& Kandrup 2002) suggests strongly that, at 
least for the case of systems in a time-independent equilibrium, such 
constructs {\em do} remain applicable. Specifically, a comparison of the same 
initial conditions evolved both in the smooth potential and in fixed (in 
space and time) $N$-body realisations of the corresponding density 
distribution reveals that, both for orbit ensembles {\em and} for individual 
orbits, $N$-body trajectories are extremely well modeled by smooth potential 
orbits perturbed by Gaussian white noise with amplitude consistent with that 
predicted by a Fokker-Planck description (Chandrasekhar 1943). Indeed, it is 
possible to extract estimates of finite time Lyapunov exponents for the smooth 
potential from such $N$-body trajectories (Kandrup \& Sideris 2003). 

This suggests that, in a first approximation, one can interpret 
orbits in $N$-body systems as exhibiting (Lichtenberg \& Lieberman 1992) 
`intrinsic diffusion', just as in the smooth potential, so that, e.g., one can 
visualise orbits passing through cantori in the fashion described by the 
`turnstile model' of 
MacKay, Meiss, and Percival (1984). In a next approximation, it would then 
seem reasonable to model discreteness effects using Fokker-Planck or Langevin 
simulations (Chandrasekhar 1943), the friction and noise associated with such 
simulations being interpreted (Lichtenberg \& Lieberman 1992) as 
a source of `extrinsic diffusion' which tends, generically, to 
accelerate phase space transport. 

\subsubsection{White noise}

The initial conditions were generated by sampling a tiny phase space hypercube of characteristic size $r{\;}{\sim}{\;}0.01$ or less. In the absence of perturbations, orbits were generated from these initial conditions by solving the Hamilton equations appropriate for motion in the potential (1), using a variable time step integrator which typically conserved energy to better than one part in $10^{5}$. Following Chandrasekhar (1943), discreteness effects, i.e., the effects of gravitational Rutherford scattering, were modeled as resulting in friction and additive white noise connected by a Fluctuation-Dissipation Theorem. The evolution of individual orbits thus entailed solving a Langevin equation (Chandrasekhar 1943, van Kampen 1981) of the form
\begin{equation}
{d^{2}x^{a}\over dt^{2}}=-{{\partial}V({\bf r})\over {\partial}x^{a}}
-{\eta}v^{a}+F^{a}, \qquad (a=x,y,z), 
\end{equation}
with ${\eta}$ a constant coefficient of dynamical friction and ${\bf F}$ a `stochastic force.' Assuming in the usual fashion that ${\bf F}$ corresponds to homogeneous Gaussian noise, its statistical properties are characterised completely by its first two moments, which take the form
\begin{displaymath}
{\langle}F_{a}(t){\rangle}=0 \qquad {\rm and}
\end{displaymath}
\begin{equation}
{\langle}F_{a}(t_{1})F_{b}(t_{2}){\rangle}=
{\delta}_{ab}\;K(t_{1}-t_{2}),
\qquad (a,b=x,y,z).
\end{equation}
The assumption that the noise be white implies further that the autocorrelation function $K$ is proportional to a Dirac delta, so that
\begin{equation}
K({\tau})=2{\eta}{\Theta}{\delta}_{D}({\tau}) .
\end{equation}
Physical considerations dictate that the `temperature' ${\Theta}{\;}{\sim}{\;}|E|$, so the simulations were all performed assuming ${\Theta}=-E$. In this case, ${\eta}$ defines the relaxation time $t_{R}$ on which the energy of an orbit evolved in an otherwise fixed potential would change significantly: $t_{R}{\;}{\equiv}{\;}{\eta}^{-1}$ The Langevin equation was solved using an algorithm developed by Griner, Strittmatter, \& Honerkamp (1988) (see also Honerkamp 1994).

Orbital data, recorded at intervals ${\delta}t=0.1t_{D}$ were binned into rectangular grids comprised of $20 \times 20$ cells so as to construct coarse-grained distribution functions $f(Z_{a},Z_{b},t)$ for pairs of phase space variables, i.e., $a{\,}{\ne}{\,}b=x,y,z,v_{x},v_{y},v_{z}$. An examination of successive `snapshots' revealed a systematic tendency for the ensemble to disperse and, eventually, to approach a nearly time-independent state. For this reason, the last $500$ snapshots, appropriate for $150.1t_{D}{\;}{\le}{\;}t{\;}{\le}{\;}200.0t_{D}$ were combined to generate a numerical representation of a coarse-grained near-invariant distribution, i.e.,
\begin{equation}
f_{niv}={1\over 500} \sum_{i=1501}^{2000} f(t_{i}).
\end{equation}
The approach of $f(t)$ towards the near-invariant $f_{niv}$ was quantified by computing an $L^{2}$ ``distance'' between $f(t)$ and $f_{niv}$ via the natural prescription (cf. Kandrup \& Mahon 1994, Merritt \& Valluri 1996, Kandrup 1998b)
\begin{equation}
$$Df(Z_{a},Z_{b},t)=
{\Biggl(} {\sum_{a}\sum_{b}|f(Z_{a},Z_{b},t)-f_{niv}(Z_{a},Z_{b})|^{2} 
\over \sum_{a}\sum_{b}|f_{niv}(Z_{a},Z_{b})|^{2} }{\Biggr)}^{1/2}. $$
\end{equation}

\subsubsection{Coloured noise}

Discreteness effects such as stellar encounters correspond to {\em instantaneous} random kicks, and can be adequately modeled as white noise. However, other discreteness effects (e.g., large-scale organised motions inside a galaxy or perturbations caused by an external environment) are better modeled as random kicks of {\em finite} dutation. Assuming that these kicks constitute a random Gaussian process, one is led once again to solve the Langevin equation (7), the only difference being that the autocorrelation function $K(t_{1}-t_{2})$ is no longer idealised as a Dirac delta (`coloured noise').

Two specific choices for the autocorrelation function were considered. One corresponds to a so-called Ornstein-Uhlenbeck process (cf. van Kampen 1981), for which $K$ decreases exponentially in time, i.e.,
\begin{equation}
K({\tau})={\alpha}{\eta}{\Theta}\,\exp(-{\alpha}|{\tau}|).
\end{equation}
This process is characterised by three parameters: ${\eta}$ and ${\Theta}$ have the same meaning as for the case of white noise, whereas the autocorrelation time 
\begin{equation}
t_{c}{\;}{\equiv}{\;}{\int_0^{\infty}{\tau}K({\tau})d{\tau}\over
\int_0^{\infty}K({\tau})d{\tau}} ={\alpha}^{-1}
\end{equation}
sets the time scale on which the random forces change appreciably. The normalisation in eq.~(12) ensures that the diffusion constant $D$ that would enter into a Fokker-Planck description,
\begin{equation}
D=\int_{-\infty}^{\infty}\,d{\tau}K({\tau})=2{\Theta}{\eta},
\end{equation}
is independent of ${\alpha}$.

The other choice (cf. Pogorelov \& Kandrup 1999) corresponds to an autocorrelation function of the form
\begin{equation}
K({\tau})={3{\alpha}{\eta}{\Theta}\over 8}\exp(-{\alpha}|{\tau}|)\,
{\Bigl(}1+{\alpha}|{\tau}|+{{\alpha}^{2}\over 3}{\tau}^{2}{\Bigr)} .
\end{equation}
For fixed ${\alpha}$ this autocorrelation function decays somewhat more slowly, the autocorrelation time now equaling $t_{c}=2/{\alpha}$. In both cases, the integrations were performed using a variant of an algorithm summarised in Pogorelov \& Kandrup (1999). White noise can be viewed as a singular ${\alpha}\to\infty$ limit of either of these two processes.

\subsection{Unperturbed Hamiltonian evolution}

As for simpler potentials (Kandrup 1998a), an initially localised ensemble of chaotic orbits tends to disperse exponentially at a rate ${\lambda}$ comparable to the typical size of the finite time Lyapunov exponents, ${\chi}$, for the ensemble. This implies, e.g., that quantities like ${\sigma}_{x}$ and ${\sigma}_{v_x}$, the dispersions in position and velocity associated with the ensemble, initially grow exponentially. More significant, however, for the problem of chaotic phase mixing, is the fact that this evolution also entails a comparatively efficient approach towards a near-invariant distribution $f_{niv}$. In most cases, this evolution towards $f_{niv}$ is approximately exponential, so that $Df(t){\;}{\propto}{\;}\exp (-{\Lambda}t)$; and, in agreement with Kandrup \& Mahon (1994) and Merritt \& Valluri (1996), one finds typically that ${\Lambda}$ is again comparable in magnitude to, albeit somewhat smaller than, a typical finite time Lyapunov exponent ${\chi}$.

\begin{figure}
\centering
\centerline{
    \epsfxsize=8cm
    \epsffile{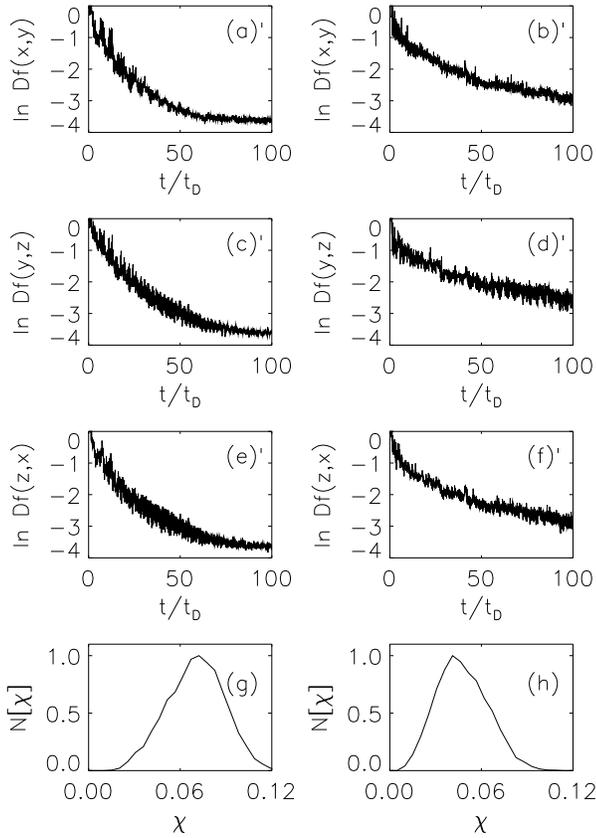}
      }
    \begin{minipage}{10cm}
    \end{minipage}
    \vskip -0.3in\hskip -0.0in
\caption{
(a) The $L^{2}$ distance $Df(x,y,t)$ between $f(x,y,t)$ and a near-invariant
$f_{niv}(x,y)$ computed for an initially localised ensemble of 1600 chaotic 
orbits in the lowest shell evolved in the triaxial Dehnen potential with 
axis ratios $c/a=1/2$ and $(a^{2}-b^{2})/(a^{2}-c^{2})=1/2$. (b) $Df(x,y,t)$
for another ensemble evolved in the same potential with the same energy.
(c) $Df(y,z,t)$ for the ensemble in (a). 
(d) $Df(y,z,t)$ for the ensemble in (b).
(e) $Df(z,x,t)$ for the ensemble in (a). 
(f) $Df(z,x,t)$ for the ensemble in (b).
(g) $N[{\chi}]$, the distribution of finite time Lyapunov exponents for the
initial conditions in (a) evolved for a time $t=200t_{D}$.
(h) $N[{\chi}]$ for the initial conditions in (b) evolved identically.
}
\label{landfig}
\end{figure}

\begin{figure}
\centering
\centerline{
    \epsfxsize=8cm
    \epsffile{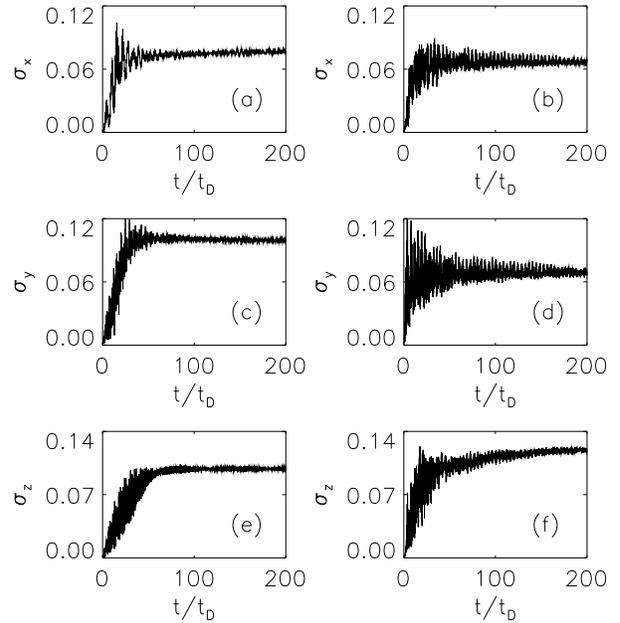}
      }
    \begin{minipage}{10cm}
    \end{minipage}
    \vskip -1.2in\hskip -0.0in
\caption{
(a) The dispersion ${\sigma}_{x}$ for the ensemble exhibited in the left panels
of FIG. 1. (b) ${\sigma}_{x}$ for the ensemble in the right panels.
(c) and (d) The corresponding ${\sigma}_{y}$'s. (e) and (f) The corresponding
${\sigma}_{z}$'s. 
}
\label{landfig}
\end{figure}

A typical example of this behaviour is illustrated in the top three left hand panels of FIGURE 8, which exhibit $Df(x,y)$, $Df(y,z)$, and $Df(z,x)$ for one initially localised ensemble of chaotic orbits in the lowest energy shell evolved in the triaxial Dehnen potential with axis ratios $c/a=1/2$ and $(a^{2}-b^{2})/(a^{2}-c^{2})=1/2$. The bottom left hand panel exhibits a distribution of finite time Lyapunov exponents for the same ensemble computed for a total interval $t=200t_{D}$. It should be noted that the saturation of $\ln Df$ at a value ${\sim}{\;}-3.5$ observed in these panels is a numerical artifact, rather than a real physical effect: Even if the data points used to generate $f(t)$ and $f_{niv}$ had been obtained by randomly sampling exactly the same continous distribution, the computed distance $Df$ between them would be nonzero because of finite number statistics. 

Occasionally, however, one finds that the approach towards $f_{niv}$ is significantly less efficient. This is, e.g., illustrated by the right hand panels of FIGURE 8, which exhibit $Df$ and the distribution of finite time Lyapunov exponents for a different ensemble of chaotic orbits evolved in the same potential and with the same energy. At very early times, the distances $Df$ computed for this ensemble are not appreciably different from those observed for the first ensemble, but after $t{\;}{\sim}{\;}15t_{D}$ or so, it is clear that the approach towards a near-invariant $f_{niv}$ has become appreciably less efficient. 

As is evident from FIGURE 9, the qualitative differences exhibited between these two ensembles are also reflected by the evolution of quantities like the dispersions ${\sigma}_{x}$, ${\sigma}_{y}$, and ${\sigma}_{z}$. At very early times the dispersions for the two different ensembles evolve in a comparatively similar fashion, but three obvious differences are evident at later times: The later time evolution for the right hand ensemble is significantly less smooth, the dispersions exhibiting high frequency variability which only damps over times $t>100t_{D}$. Moreover, it is apparent that, even after a time as long as $t=200t_{D}$, ${\sigma}_{z}$ is exhibiting a systematic secular evolution. Both these features would suggest that the ensemble is comparatively inefficient in achieving a near-invariant distribution, consistent with what one infers from FIGURE 8. The third point is that the asymptotic values towards which the ${\sigma}_{i}$ converge are different for the two ensembles. This reflects the fact that the near-invariant $f_{niv}$'s for the two ensembles differ. It would appear that, at early times, the two ensembles are restricted to different phase space rehions although,at late times, they will presumably converge towards the same invariant $f_{inv}$.

\subsection[]{Evolution including discreteness effects\\* modeled as white noise}

Discreteness effects, modeled as friction and white noise, can accelerate the approach towards an invariant or near-invariant distribution in two relatively distinct fashions. On the one hand, such irregularities can accelerate phase space transport between different chaotic phase space regions which, presumably, are connected by cantori or the Arnold web. On the other, they can significantly increase the efficiency with which orbits disperse within a single chaotic phase space region so that, e.g., the rate ${\Lambda}$ associated with the initial approach towards a near-invariant distribution becomes larger.

That discreteness effects can accelerate phase space transport through topological obstructions is a fact that has been recognised previously, both in the context of the triaxial Dehnen potentials (Siopis \& Kandrup 2000) and for simpler two and three degree of freedom systems (cf. Kandrup, Pogorelov \& Sideris 1999). As first discussed in Pogorelov \& Kandrup (1999), this can be understood as reflecting the fact that weak irregularities wiggle the orbits, thus assisting them in finding phase space holes.

That discreteness effects can also accelerate phase space transport within a given phase space region does not seem to have been recognised previously. The fact that introducing weak irregularities can increase ${\Lambda}$ is significant both for its physical implications -- which will be discussed more carefully below -- and for what it may suggest about the physical mechanism responsible for chaotic phase mixing. It is hardly surprising that an initially localised ensemble of chaotic orbits evolved into the future will disperse at a rate ${\lambda}$ comparable to a typical finite time Lyapunov exponent ${\chi}$ for the orbits in the ensemble. However, it is not completely obvious how, if at all, the rate ${\Lambda}$ associated with the exponential approach towards a near-invariant distribution should correlate with ${\chi}$. It is thus interesting, and significant, that, for a variety of different potentials (cf. Kandrup \& Mahon 1994, Merritt \& Valluri 1996), in the absence of perturbations there seems to be a strong correlation between ${\chi}$ and ${\Lambda}$, although the correlation is not completely linear. Overall, ${\Lambda}$ seems to be a factor of 2 or 3 smaller than ${\chi}$.

Except for the very largest perturbations, ${\eta}>10^{-2.5}$ or so, friction and noise do {\it not} affect the exponential rate at which orbits in an ensemble disperse, although they {\it do} have a subexponential effect. If, e.g., one selects an ensemble that samples a very tiny region (or, as an extreme case, simply tracks multiple noisy realisations of the same initial condition), the dispersions will grow as 
\begin{equation}
{\sigma}_{x}{\;}{\propto}{\;}({\Theta}{\eta})^{1/2}\exp({\chi}t)
\end{equation}
(cf. Habib, Kandrup, \& Mahon 1997). The irregularities facilitate the spreading of the ensemble, but they do not change the overall rate! By contrast, friction and noise {\it do} increase the rate ${\Lambda}$ at which the orbit ensemble evolves towards a near-invariant distribution.

This increase is evident visually from FIGURES 10 and 11, which exhibit $Df(x,y,t)$ for the same two sets of initial conditions used to generate FIGURES 8 and 9, now evolved allowing for friction and white noise with variable $t_{R}$ between $10^{4}t_{D}$ and $10^{6.5}t_{D}$. Best fit values of ${\Lambda}$ for the initial interval $0{\;}{\le}{\;}t{\;}{\le}{\;}6t_{D}$, both for these and other values of $t_{R}$, are exhibited in the top two panels of FIGURE 12. In each case, the computed value averages over the fifteen possible choices of $f(Z_{a},Z_{b})$, and the formal error bars correspond to the standard deviations associated with the mean. The dashed lines correspond to the best fit value of ${\Lambda}$ for simulations without any friction or noise. 

Although a useful probe of the early stages of the evolution towards a near-invariant distribution, ${\Lambda}$ misses the clear differences between the two ensembles that are obvious visually in FIGURE 8. An alternative probe, more sensitive to somewhat later time evolution, is the time ${\tau}$ required before $Df(t)$ becomes smaller than some fiducial value. The bottom two panels of FIGURE 12 exhibit ${\tau}_{.05}$, the time required for $Df(t)$, which starts from an initial value $Df=1$, to decrease to a value $Df(t)=0.05$. Perhaps the most striking thing about FIGURE 12 is the fact that even very weak friction and noise, corresponding to $t_{R}>10^{6}t_{D}$, is enough to dramatically accelerate the approach of the second ensemble towards a near-invariant distribution. The best fit values of ${\Lambda}$ in the absence of noise and for $t_{R}=10^{6.5}t_{D}$ are virtually identical, but the noisy simulation is characterised by a value of ${\tau}_{.05}$ that is less than half as large as the value assumed in the absence of noise! 

Another interesting feature, again evident from FIGURE 12, is that ${\Lambda}$ and ${\tau}_{.05}$ both exhibit a roughly logarithmic dependence on ${\eta}$ and $t_{R}$. This logarithmic dependence, observed also when probing the effects of friction and noise on phase space transport through cantori in two-degree-of-freedom systems (Pogorelov \& Kandrup 1999), implies that the effects of the perturbations only turn on gradually. There is no critical threshhold amplitude above which the perturbations immediately become important. Equation (12) facilitates at least a heuristic explanation of why the time ${\tau}$ required to converge towards $f_{niv}$ scales logarithmically in ${\eta}$ or $t_{R}$. If one assumes, simplistically, that ${\tau}$ should scale as the time required for the ensemble to expand to a size comparable to the accessible phase space region, ${\tau}$ corresponds to a time when the configuration space dispersions have assumed some fiducial value. Assuming, however, that this be true, eq. (16) implies that ${\tau}{\;}{\propto}{\;}{\rm const} + \ln t_{R}$.

The effects of noise on orbits inside a given phase space region also resemble the effects of noise on diffusion between different phase space regions in one other important respect: the details seem largely unimportant. Turning off the friction but retaining the noise, or making the white noise multiplicative (i.e., state-dependent), so that ${\eta}$ is a nontrivial function of ${\bf x}$ and/or ${\bf v}$, seems largely irrelevant. 

In any event, what is apparent, e.g., from FIGURE 12, is that even comparatively low amplitude perturbations can dramatically accelerate chaotic phase mixing within a given phase space region. Friction and noise corresponding to a relaxation time as long as $t_{R}{\;}{\sim}{\;}10^{5}t_{D}$ can increase the rate of chaotic phase mixing by ${\sim}{\;}50\%$; friction corresponding to $t_{R}{\;}{\sim}{\;}10^{4}t_{D}$ can increase ${\Lambda}$ by a factor of two. What this suggests is that, even in settings where diffusion through cantori is comparatively unimportant, friction and noise can play an important role in enhancing the overall efficacy with which orbits disperse and, presumably, the rate at which a configuration displaced from equilibrium can readjust towards a new equilibrium.

\begin{figure}
\centering
\centerline{
    \epsfxsize=8cm
    \epsffile{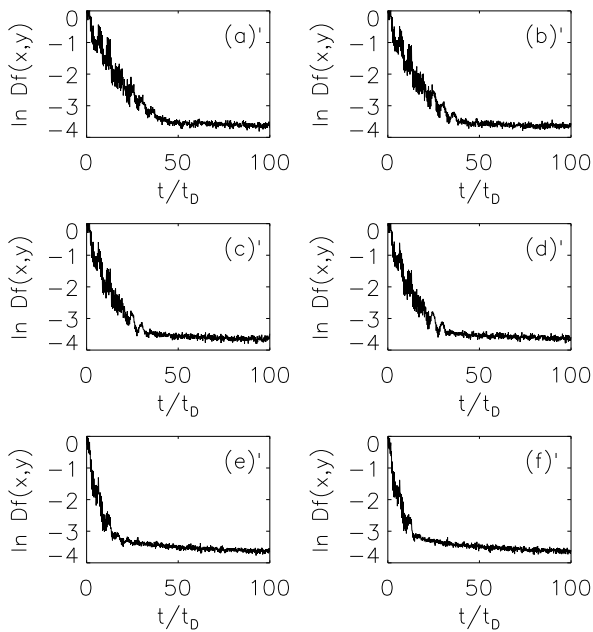}
      }
    \begin{minipage}{10cm}
    \end{minipage}
    \vskip -1.2in\hskip -0.0in
\caption{
The $L^{2}$ distance $Df(x,y,t)$ between $f(x,y,t)$ and a near-invariant
$f_{inv}(x,y)$ computed for the first ensemble in FIG. 1, now perturbed by 
friction and white noise with ${\Theta}=-E$ and variable ${\eta}=t_{R}^{-1}$. 
(a) $t_{R}=10^{6.5}t_{D}$. (b) $t_{R}=10^{6}t_{D}$. (c) $t_{R}=10^{5.5}t_{D}$.
(d) $t_{R}=10^{5}t_{D}$. (e) $t_{R}=10^{4.5}t_{D}$. (f) $t_{R}=10^{4}t_{D}$.
}
\label{landfig}
\end{figure}

\begin{figure}
\centering
\centerline{
    \epsfxsize=8cm
    \epsffile{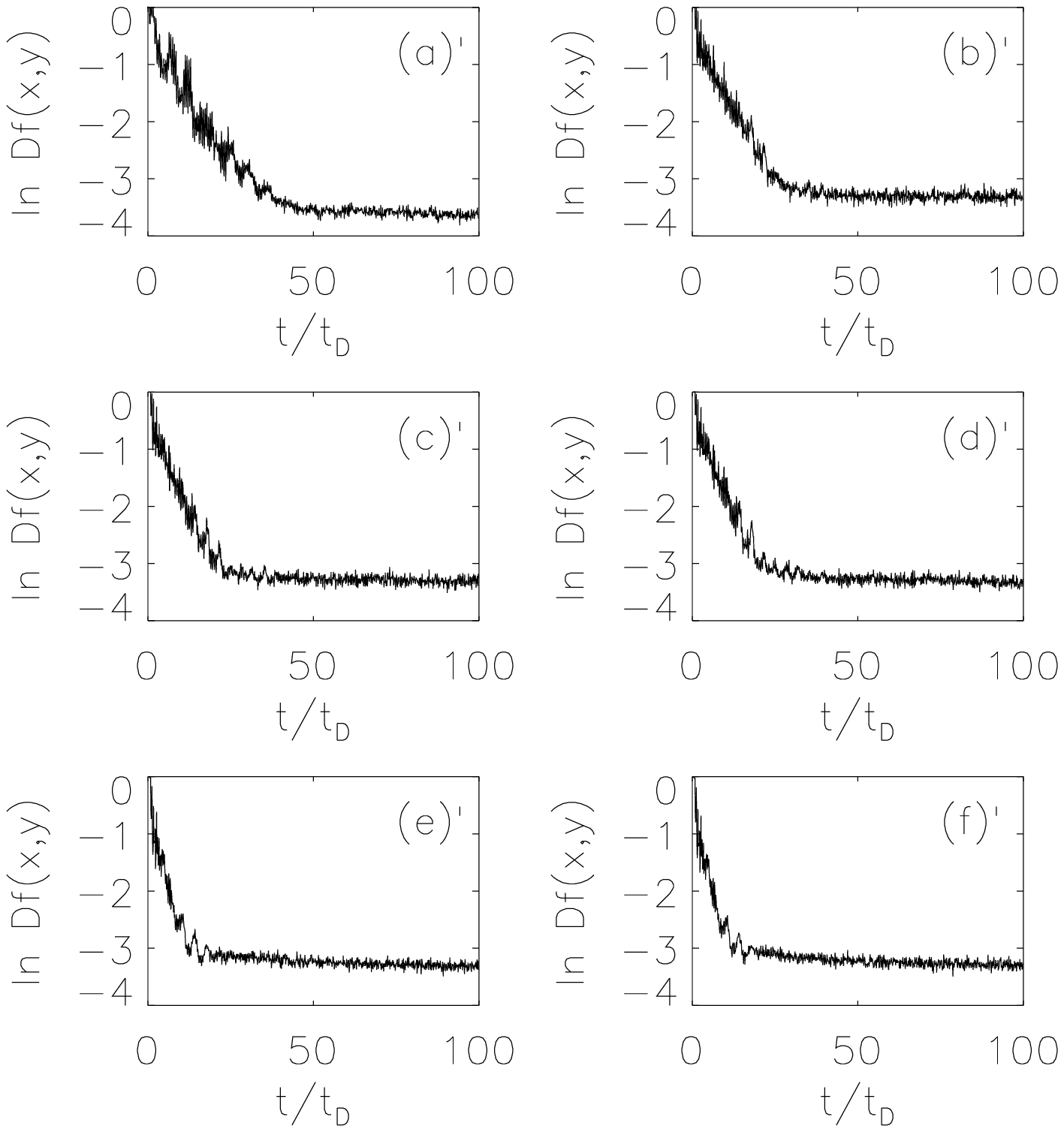}
      }
    \begin{minipage}{10cm}
    \end{minipage}
    \vskip -1.2in\hskip -0.0in
\caption{
The $L^{2}$ distance $Df(x,y,t)$ between $f(x,y,t)$ and a near-invariant
$f_{inv}(x,y)$ computed for the second ensemble in FIG. 1, now perturbed by 
friction and white noise with ${\Theta}=-E$ and variable ${\eta}$. 
(a) $t_{R}=10^{6.5}t_{D}$. (b) $t_{R}=10^{6}t_{D}$. (c) $t_{R}=10^{5.5}t_{D}$.
(d) $t_{R}=10^{5}t_{D}$. (e) $t_{R}=10^{4.5}t_{D}$. (f) $t_{R}=10^{4}t_{D}$.
}
\label{landfig}
\end{figure}

\begin{figure}
\centering
\centerline{
    \epsfxsize=8cm
    \epsffile{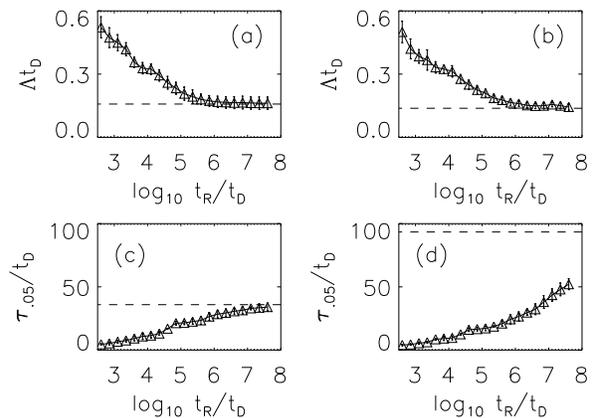}
      }
    \begin{minipage}{10cm}
    \end{minipage}
    \vskip -2.4in\hskip -0.0in
\caption{
(a) ${\Lambda}$, the rate of approach towards a near-invariant distribution
for the first ensemble of initial conditions in FIG. 1, allowing for friction 
and white noise with variable $t_{R}={\eta}^{-1}$. (b) The same for the second 
ensemble.
(c) ${\tau}_{.05}$, the time required for the coarse-grained $f$ associated
with the first ensemble to evolve towards the near-invariant $f_{niv}$ at the
5\% level. (d) The same for the second ensemble.
In each panel, the dashed line represents the value of ${\Lambda}$ or 
${\tau}_{.05}$ observed in the absence of friction and noise. 
}
\label{landfig}
\end{figure}

\subsection[]{Evolution including perturbations modeled as\\* coloured noise}

Just as for white noise, it was found that coloured noise can significantly accelerate phase space transport both by facilitating transport along the Arnold web and by enhancing diffusion within a single chaotic phase space region. In particular, one discovers once again that the initial evolution towards a near-invariant distribution $f_{niv}$ is typically well fit by an exponential, at least at early times, and, for fixed ${\alpha}$ or $t_{c}$, that the rate ${\Lambda}$ associated with this exponential approach again scales logarithmically with ${\eta}$ or $t_{R}$.

As was observed also for other potentials (cf. Pogorelov \& Kandrup 1999, Kandrup, Pogorelov \& Sideris 2000), the response of an orbit ensemble seems insensitive to most details. For fixed $t_{c}$ the two types of noise that were considered had very similar effects; and the presence or absence of dynamical friction also proved largely irrelevant. All that really seems to matter are the values of $t_{c}$, which sets the time scale on which the noise changes appreciably, and $t_{R}$, which probes the overall amplitude of the noise. In this sense, the results of these experiments were all consistent with the interpretation presented by Pogorelov \& Kandrup (1999), namely that noise acts through a resonant coupling with an orbit. 

The spectral density, $S({\omega})$, given as the Fourier transform of the autocorrelation function $K({\tau})$, characterises the degree to which the noise has significant power at different frequencies ${\omega}$. The crucial point then is that the noise will have a significant effect on an orbit if and only if $S$ has significant power at frequencies ${\omega}$ corresponding to the frequencies ${\Omega}{\;}{\sim}{\;}t_{D}^{-1}$ for which the orbit itself has significant power. It follows that, for autocorrelation times $t_{c}{\;}{\gg}{\;}t_{D}$, coloured noise becomes comparatively ineffectual as a source of accelerated phase space transport.

\begin{figure}
\centering
\centerline{
    \epsfxsize=8cm
    \epsffile{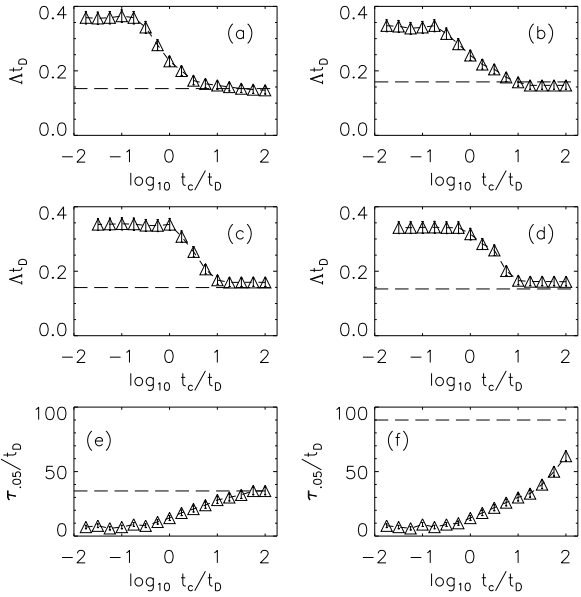}
      }
    \begin{minipage}{10cm}
    \end{minipage}
    \vskip -0.in\hskip -0.0in
\caption{
(a) ${\Lambda}$, the rate of approach towards a near-invariant distribution
for the first ensemble of initial conditions in FIG. 1, allowing for friction 
and Ornstein-Uhlenbeck coloured noise with $t_{R}=4000t_{D}$ and variable 
$t_{c}$. The dashed line corresponds to the rate ${\Lambda}$ for unperturbed
evolution.
(b) The same for the second ensemble. (c) and (d) The same as (a)
and (b), now allowing for coloured noise with autocorrelation function given
by eq. (16).
(e) ${\tau}_{.05}$, the time required for the coarse-grained $f$ associated
with the first ensemble to evolve in the presence of Ornstein-Uhlenbeck noise
towards the near-invariant $f_{niv}$ at the
5\% level. (f) The same for the second ensemble.
}
\label{landfig}
\end{figure}

Examples of the effects of varying $t_{c}$ for fixed $t_{R}$ and fixed form of the coloured noise are illustrated in FIGURE 13, which was generated once again from the ensembles of initial conditions used to generate FIGURE 8. The top two panels exhibit the effects of Ornstein-Uhlenbeck noise with $t_{R}=4000t_{D}$ on the convergence rate ${\Lambda}$; the middle two exhibit the effects of noise with autocorrelation function given by eq. (16). The bottom two panels exhibit ${\tau}_{.05}$ for the ensembles used to generate panels (a) and (b). In each case, it is clear that, for $t_{c}{\;}{\ll}{\;}t_{D}$, the value of $t_{c}$ is essentially irrelevant and ${\Lambda}$ assumes the value appropriate for white noise. Alternatively, for $t_{c}{\;}{\gg}{\;}t_{D}$ the noise has a comparatively minimal effect. For values of $t_{c}$ comparable to or somewhat larger than $t_{D}$, ${\Lambda}$ is a smoothly decreasing function which exhibits a roughly logarithmic dependence on $t_{c}$.

That ${\Lambda}$ only begins decreasing significantly for somewhat larger values of $t_{c}$ for the fourth order noise with autocorrelation function given by eq.~(16) reflects the fact that, for fixed $t_{R}$, the autocorrelation function decreases somewhat more slowly with $t_{c}$ than for the case of Ornstein-Uhlenbeck noise. This is, e.g. illustrated in FIGURE 14, which exhibits graphically the autocorrelation functions (12) and (15), in each case allowing for an autocorrelation time $t_{c}=1.0$ and a diffusion constant $D=1000$. ${\Lambda}(t_{c})$ should not begin to decrease until $t_{c}$ becomes sufficiently large that there is an appreciable decrease in power for frequencies ${\sim}{\;}t_{D}^{-1}$ relative to $t_{c}=0$.

\begin{figure}
\centering
\centerline{
    \epsfxsize=8cm
    \epsffile{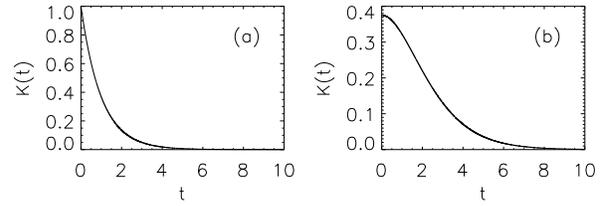}
      }
    \begin{minipage}{10cm}
    \end{minipage}
    \vskip -1.25in\hskip -0.0in
\caption{
(a) The autocorrrelation function $K(t)$ for an Ornstein-Uhlenbeck
process, plotted for $t_{c}=1.0$ and $D=1000$. (b) The same for the
autocorrelation function given by eq. (16).
}
\label{landfig}
\end{figure}

Interestingly, however, even very low frequency noise, with $t_{c}{\;}{\gg}{\;}t_{D}$, {\it can} have an appreciable effect on quantities like ${\tau}_{.05}$. Even though such low frequency noise does not dramatically accelerate the initial exponential approach towards a near-invariant distribution, it {\it does} decrease significantly the time required for the second ensemble to approach a near-invariant distribution. This fact, illustrated in panels (e) and (f), reflects the fact that, even though the noise does not significantly impact the ``average'' behaviour of quantities like the dispersion ${\sigma}_{x}$, it does suppress the large amplitude fluctations which are evident in the right hand panels of FIGURE 9.

\section{SUMMARY}

The aim of this paper, along with Paper I, has been to investigate the role of chaotic processes in the triaxial Dehnen potential, which has been considered a realistic approximation for at least the central regions of many elliptical galaxies and, possibly, also for some spiral galaxy bulges. Particular emphasis was placed on the way in which these processes are affected by the presence of internal and/or external perturbations, modeled as periodic driving (Paper I) and/or dynamical friction plus white or coloured noise. Such perturbations should in fact be operative in most real galaxies.

The major findings can be summarised as follows:
\begin{enumerate}

\item In the presence of a central density cusp ($\gamma > 0$) and/or a central point singularity (black hole), the fraction $f$ of `strongly chaotic' orbits generally increases with increasing $\gamma$, with decreasing distance from the centre, with increasing deviation from sphericality towards axisymmetry, and with increasing deviation from axisymmetry towards triaxiality. Typical values for $f$ in triaxial configurations range between 30 and 70 percent. For a given deviation from sphericality towards axisymmetry, $f$ is higher when the deviation is prolate than when it is oblate.

\item The fact that there are unquestionably regular orbits passing very close to the centre suggests that chaos is probably triggered by appropriate resonance overlaps, which are stronger in steeper cusps, rather than `close encounters' with the central cusp.

\item In the absence of a central density cusp ($\gamma = 0$) {\em and} of a central black hole, the fraction of strongly chaotic orbits \emph{increases} with distance from the centre. This is interpreted as evidence that the central cusp plays a dominant role in generating chaos for $\gamma > 0$.

\item The exponential instability of the strongly chaotic orbits, as measured by the value of their maximal Lyapunov exponent, is characterised by a timescale of order a few dynamical times. However, unlike many other two- and three-dimensional potentials, chaotic orbits in the triaxial Dehnen potential tend to be extremely `sticky,' even for integration times of $20,000\;t_D$ or longer.

\item The presence of a central black hole as large as 1\% of the total mass causes a relatively small increase in the fraction of strongly chaotic orbits, mainly near the centre. However, the values of maximal Lyapunov exponents can increase substantially. 

\item Analogous results were also obtained for a toy model consisting of an anisotropic oscillator with a softened Plummer sphere (supermassive black hole) superimposed in the centre (Kandrup \& Sideris 2002). This suggests that they may be generic to nonaxisymmetric systems with central density cusps and/or central black holes. 

\item The evolution of an initially localised ensemble of chaotic orbits is typically characterised by three time scales, two comparatively short and the third much longer. The initial, shortest time scale evolution corresponds to an exponential divergence at a rate ${\lambda}$ comparable to a typical value of the largest finite-time Lyapunov exponent ${\chi}$ for the orbits in the ensemble. This is followed by an exponential approach towards a near-invariant distribution $f_{niv}$ at a somewhat slower, but still comparable, rate $\Lambda \la \lambda \approx \chi$. This comparatively rapid evolution is then followed by a much slower evolution as the orbits diffuse along Arnold webs so as to, eventually, sample the true equilibrium. Since ${\Lambda}^{-1}$ is typically comparable to a typical dynamical time, the first two stages correspond to timescales $\tau\;{\sim}{\;}t_{D}$; the time scale for the third stage is typically much longer, $\tau\;{\gg}{\;}t_{D}$.

\item All three of these timescales can be shortened considerably by allowing for discreteness effects associated with two-body relaxation and modeled as a superposition of friction and additive or multiplicative homogeneous Gaussian white noise in the context of a Langevin, or Fokker-Planck, description. 

\item Coloured noise, corresponding to finite-duration random kicks characterised by an autocorrelation time $t_c$, can have the same effects as white noise, provided that $t_c \la t_D$. 

\end{enumerate}

\section{IMPLICATIONS FOR REAL GALAXIES}

In assessing the implications of these results for real galaxies, it is useful to differentiate between three relatively distinct scenaria, which are discussed in more detail in the following sections:

\begin{enumerate}

\item {\em No external perturbations:} Relatively isolated `field' galaxies, for which the time-independent Hamiltonian description is largely adequate. However, internal discreteness effects are still at work, and can affect the efficiency with which such a galaxy can reach a (quasi-)equilibrium. 

\item {\em Weak external perturbations:} Galaxies with a small number of satellite/companion objects and/or belonging to a moderately dense cluster of galaxies, where the (internal) galactic bulk potential still remains approximately time independent. Internal perturbations can still be important, as in isolated galaxies.

\item {\em Strong perturbations:} The later stages of violent interactions where the time-independent approximation completely fails. Examples of such interactions include galaxy mergers and frequent high-speed galaxy encounters (`galaxy harassment'). In such cases, a `violent relaxation' approximation is more appropriate.

\end{enumerate}

This discussion does not include the effects of a dissipative component (cold gas), so it is perhaps most appropriate for elliptical galaxies and for certain spiral bulges.

\subsection{Isolated galaxies (no external perturbations)}

Important discreteness effects inside a galaxy include:

\begin{enumerate}

\item Stellar (two-body) encounters, where the distance between objects is very small compared with the size of the galaxy. These can be modeled adequately as instantaneous ($t_c \ll t_D$; cf. Eq.~(17) in Section~5.2) random kicks, i.e., as friction and white noise characterised by a relaxation time $t_R = \eta^{-1}$ and a temperature $\Theta = -E$. FIGURES 10 and 11 exhibit the effects of white noise for a few values of $E$ and several values of $t_R$.

\item Large-scale organised motions and encounters with massive, extended objects such as giant molecular clouds and star clusters. For these somewhat larger objects, $t_c$ is no longer completely negligible compared with $t_D$. However, to the extent that $t_c/t_D \la 0.2$ or so, the results of Section~3.4 indicate that white noise should still be adequate. Whether or not this is the case can depend on details such as the velocity field in the galaxy. For instance, $t_c/t_D$ can be $\ll 1$ in `thermally' supported objects, such as many elliptical galaxies, where the duration of encounters can be short. However, in locales where the velocity dispersions are small and motion is highly organised, such as in disks of spiral galaxies, it may be that $t_c \sim t_D$ or even $t_c \gg t_D$ (consider, e.g., the long `encounter' of a star with a molecular cloud, both moving along near-circular orbits of similar radii in the disk of a spiral galaxy). Such encounters can be seen as random kicks of finite duration which could be modeled as friction and coloured noise with appropriate autocorrelation times $t_c$.

Although the presence and properties of this coloured noise depend on the details of the internal kinematics of the galaxy, its dynamical consequences should be insensitive to these details. As shown in Section~3.4, the effects of coloured noise are much the same as for white noise when $t_c \la t_D$; and since, for fixed amplitude, coloured noise has a much weaker effect for $t_c \gg t_D$, it should prove largely irrelevant for isolated galaxies which do not experience the effects of very massive extended objects.

\end{enumerate}

What are the dynamical implications of these internal irregularities? Conventional wisdom (cf. e.g. Binney \& Tremaine 1987) assumes that the dynamics of isolated galaxies is governed solely by the bulk potential of the constituent stars because discreteness effects such as two-body interactions do not alter appreciably the values of the integrals of motion (such as the energy $E$) over the course of a Hubble time (cf. Chandrasekhar 1941). This treatment should, indeed, be valid for integrable potentials where the motion is completely regular, as well as for hyperbolic nonintegrable potentials where motion is completely ergodic.

However, it may {\em not} be appropriate to neglect discreteness effects in potentials admitting a large measure of both regular and chaotic orbits, such as the triaxial Dehnen potential and the majority of nonspherical potentials used in galactic dynamics, even though the values of the integrals of the motion still do remain approximately constant. Chaotic orbits in such potentials are often `sticky,' i.e., when integrated over some finite time, they can spend a large fraction of that time `trapped' inside cantori or within distinct phase space regions separated by the Arnold web. However, as explained in the preceding sections, discreteness effects can dramatically accelerate the rate at which sticky orbits become `untrapped' and approach their respective (near-)\linebreak invariant distributions.

These considerations can be important in at least two ways. First, discreteness effects, by acting as a continuous `microharassment' that violates Liouville's theorem, tend to smooth away substructures of phase space which correspond to regions of extreme instability, such as homoclinic points or high-order cantori. This can be of relevance when using fixed potentials as realistic galactic models. A study of the motion near unstable regions in such potentials should probably use realistic levels of noise during orbit integrations, corresponding to the graininess expected in the actual galactic potential, in order to improve the robustness against {\em structural} perturbations in the form and in the Hamiltonian character of the potential.

A second consideration involves the construction of self-consistent equilibrium models of galaxies using orbital superposition methods such as Schwarzschild's method (Schwarzschild, 1979) and more recent variants thereof. The objective of such methods is to find a weighted superposition of orbits that reproduces the mass distribution generating the potential in which the orbits were evolved. In order for the model to be a true equilibrium, each superimposed (`library') orbit should correspond to a time-independent building block, i.e., to a time-averaged (and hence space-averaged) approximation of the invariant distribution formed by galaxy orbits which share a particular combination of values for the applicable integral(s) of the motion. However, a library orbit which remains sticky throughout much of its numerical integration time may not be a good time-independent building block since it does not sample uniformly its invariant measure. \emph{This remains true even if there are `real' sticky orbits, in the `real' galaxy, which never become `unstuck' over a Hubble time!} It is, therefore, desirable to integrate library orbits using noise with an amplitude chosen so that it accelerates phase space transport as much as possible, without adversely affecting the properties of the potential. In practice, this usually means that one can safely use a noise amplitude corresponding to the expected relaxation time of the galaxy or the accuracy of the numerical integrator, whichever is greater. If this level of noise is not enough to disentangle the sticky orbits, then one could experiment with further increasing the amplitude, as long as the salient properties of the potential remain unaffected, and/or increasing the integration time.

It should be noted at this point that it is not always realistic to expect that the invariant measure can be sampled well for all potentials using integration times which are not excessively long. The extreme stickiness exhibited, for example, by the cuspy triaxial potentials studied in this paper and in Paper I, makes it very hard to achieve a good numerical approximation of the invariant measure. In a strict sense, this would mean that it is not possible to construct equilibria of such potentials using Schwarzschild-like methods. However, it may also be the case that `partially mixed' near-equilibria can be realistic approximations over a Hubble time, especially when the partial mixing refers mostly to the outer parts of the galaxy, where $t_D$ is very long. One cannot properly answer this question without actually constructing Schwarzschild models for these potentials.

Finally, a legitimate question from a numerical standpoint concerns the properties of the noise that should be used (other than its amplitude), and whether one should use white or coloured noise or some combination of the two. The good news here, as explained in the preceding sections, is that the details of the noise are largely irrelevant, as long as the autocorrelation time $t_c \la t_D$. Since white noise can be seen as coloured noise with $t_c \rightarrow 0$, and considering that coloured noise is harder to code and requires considerably more CPU cycles to compute, it follows that simply using white noise of the appropriate amplitude is completely adequate in most cases. It is interesting to note that the `numerical noise' associated with the integration of orbits could, perhaps, play the role of a white noise generator, under appropriate circumstances; however, this is a question that would require a separate investigation.

\subsection{Weak external perturbations}

The discussion in the preceding subsection concerning the consequences of discreteness effects in isolated galaxies, obviously applies fully also in the case of a galaxy surrounded by other objects. However, these objects act as additional sources of time-dependent perturbations, and their effects are inversigated here.

In a first approximation, the effects of a companion and other nearby objects can be modeled in the spirit of Chandrasekhar's (1941) `nearest neighbour approximation,' as justified by Chandrasekhar \& von Neuman (1942) (see also Kandrup 1981), which associates the random gravitational force acting on any given object with one or two particularly proximate neighbouring objects. Let $v$ denote the typical speed of stars in the original galaxy and $u$ the typical speed with which other nearby galaxies are moving with respect to that galaxy. Similarly, let $r$ be a measure of the physical size of the original galaxy and $d$ the typical separation between galaxies. And finally, let $M$ denote the mass of the original galaxy and $m$ the mass of a typical nearby galaxy.

The natural time scale on which the external forces acting on the original galaxy change significantly should be smaller than, or comparable to, the time required for a nearby galaxy to travel a distance ${\sim}{\;}d$, so that $t_{c}$ should be no larger than
\begin{equation}
t_{c}{\;}{\sim}{\;}{d\over u}={\biggl(}{d\over r}{\biggr)}
{\biggl(}{v\over u}{\biggr)}t_{D}
\end{equation}
where $t_{D}{\;}{\sim}{\;}r/v$ denotes a characteristic dynamical time for the original galaxy. The force per unit mass associated with such a nearby galaxy will have a typical size
\begin{equation}
F{\;}{\sim}{\;}{Gm\over d^{2}}={GM\over r^{2}}{\biggl(}{m\over M}{\biggr)}
{\biggl(}{r\over d}{\biggr)}^{2}.
\end{equation}
On dimensional grounds, $t_{R}$ satisfies (cf. Eq. (11))
\begin{equation}
F^{2}t_{c}{\;}{\sim}{\;}v^{2}t_{R}^{-1},
\end{equation}
so that 
\begin{equation}
t_{R}{\;}{\sim}{\;}{v^{3}r^{3}\over G^{2}M^{2}}{\biggl(}{M\over m}{\biggr)}^{2}
{\biggl(}{d\over r}{\biggr)}^{3}{\biggl(}{u\over v}{\biggr)}.
\end{equation}
However, an application of the the Virial Theorem, 
$GM/r{\;}{\sim}{\;}v^{2}$, then implies that 
\begin{equation}
t_{R}{\;}{\sim}{\;}{\biggl(}{M\over m}{\biggr)}^{2}
{\biggl(}{d\over r}{\biggr)}^{3}{\biggl(}{u\over v}{\biggr)}\;t_{D}.
\end{equation}

The presence of satellite/companion objects or of a nearby/surrounding cluster of galaxies will typically incur comparatively low-amplitude irregularities in the bulk potential associated, e.g., with a close encounter that has displaced the galaxy from a near-equilibrium. If such a systematic time-dependence is present, $t_{c}$ could become appreciable. In this case, the lowest frequency contributions might be expected to correspond to one or two quasinormal modes characterised by frequencies ${\omega}{\;}{\sim}{\;}t_{D}^{-1}$, but one might also expect a larger collection of higher frequency modes which, in a first approximation, could be modeled as inducing random irregularities with characteristic time scale $t_{c}$ comparable to, but somewhat smaller than, $t_{D}$.

\begin{figure}
\centering
\centerline{
    \epsfxsize=8cm
    \epsffile{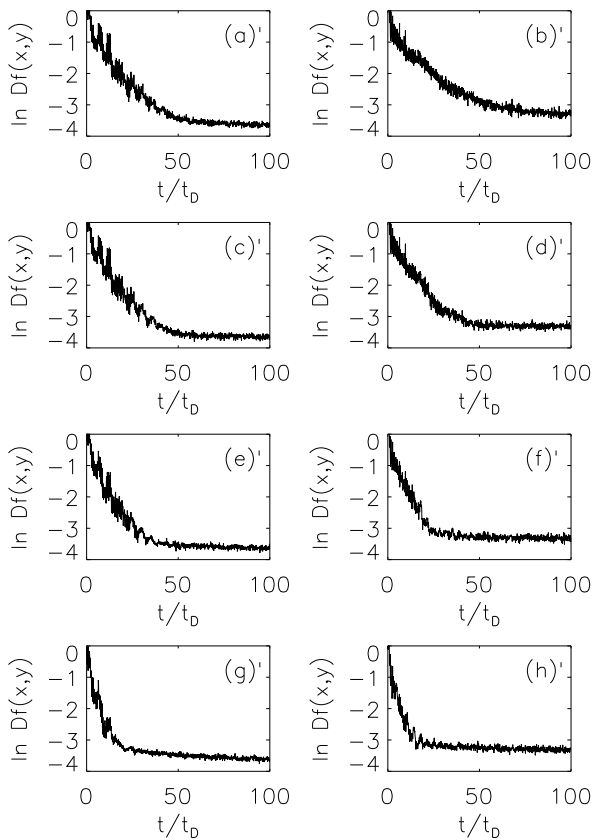}
      }
    \begin{minipage}{10cm}
    \end{minipage}
    \vskip -1.25in\hskip -0.0in
\caption{ 
 (a) The $L^{2}$ distance $Df(x,y,t)$ between $f(x,y,t)$ and a near-invariant
$f_{niv}(x,y)$ for the first ensemble in FIG. 1, now perturbed by friction
and Ornstein-Uhlenbeck colored noise with ${\Theta}=-E$, 
$t_{R}=10^{6}t_{D}$, and $t_{c}=4t_{D}$. 
(b) The same for the second ensemble in FIG. 1.
(c) The same as (a) but with $t_{R}=10^{5}t_{D}$.
(d) The same as (b) but with $t_{R}=10^{5}t_{D}$.
(e) The same as (a) but with $t_{R}=10^{4}t_{D}$.
(f) The same as (b) but with $t_{R}=10^{4}t_{D}$.
(g) The same as (a) but with $t_{R}=10^{3}t_{D}$.
(h) The same as (b) but with $t_{R}=10^{3}t_{D}$
}
\label{landfig}
\end{figure}

\begin{figure}
\centering
\centerline{
    \epsfxsize=8cm
    \epsffile{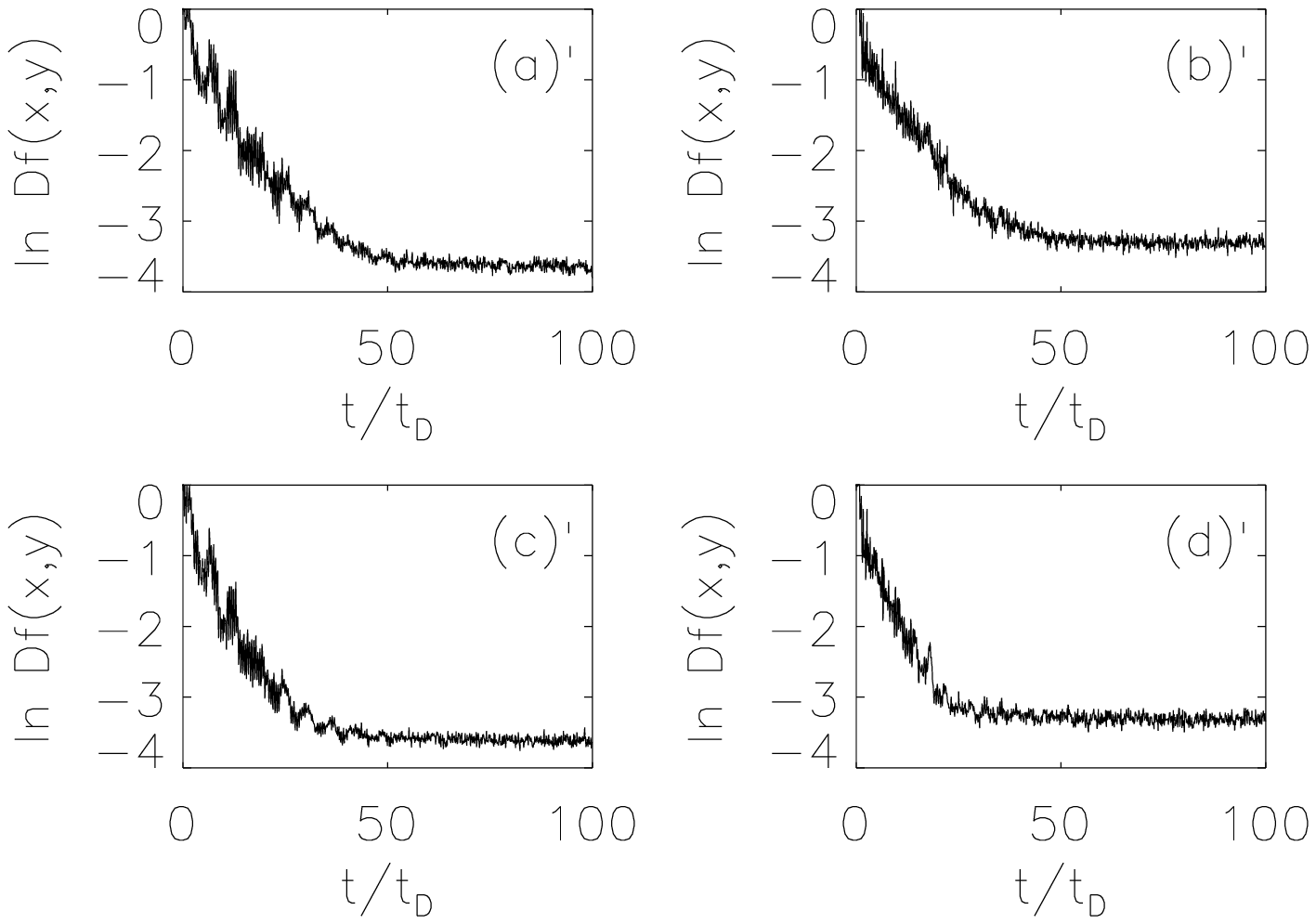}
      }
    \begin{minipage}{10cm}
    \end{minipage}
    \vskip -2.4in\hskip -0.0in
\caption{ 
 (a) The $L^{2}$ distance $Df(x,y,t)$ between $f(x,y,t)$ and a near-invariant
$f_{niv}(x,y)$ for the first ensemble in FIG. 1, now perturbed by friction
and Ornstein-Uhlenbeck colored noise with ${\Theta}=-E$, 
$t_{R}=10^{4}t_{D}$, and $t_{c}=12t_{D}$. 
(b) The same for the second ensemble in FIG. 1.
(c) The same as (a) but with $t_{R}=10^{3}t_{D}$
(c) The same as (d) but with $t_{R}=10^{3}t_{D}$
}
\label{landfig}
\end{figure}

\begin{figure}
\centering
\centerline{
    \epsfxsize=8cm
    \epsffile{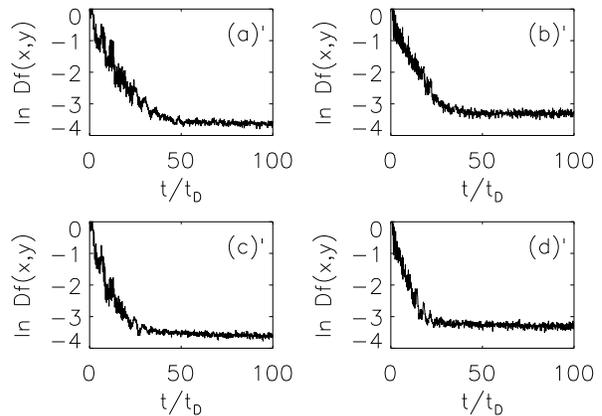}
      }
    \begin{minipage}{10cm}
    \end{minipage}
    \vskip -2.4in\hskip -0.0in
\caption{ 
 (a) The $L^{2}$ distance $Df(x,y,t)$ between $f(x,y,t)$ and a near-invariant
$f_{niv}(x,y)$ for the first ensemble in FIG. 1, now perturbed by friction
and Ornstein-Uhlenbeck colored noise with ${\Theta}=-E$, 
$t_{R}=10^{6}t_{D}$, and $t_{c}=0.4t_{D}$. 
(b) The same for the second ensemble in FIG. 1.
(c) The same as (a) but with $t_{R}=10^{5}t_{D}$.
(d) The same as (b) but with $t_{R}=10^{5}t_{D}$.
}
\label{landfig}
\end{figure}

Simple, but not unrealistic, models allowing for the effects of multiple satellite galaxies and other companion objects might entail the choices $d{\;}{\sim}{\;}4r$ and $u{\;}{\sim}{\;}v$, for which $t_{c}{\;}{\sim}{\;}4t_{D}$ and $t_{R}{\;}{\sim}{\;}64(M/m)^{2}t_{D}$. Similarly, allowing for interactions between galaxies of comparable size in a dense environment such as provided by the Coma Cluster might entail $d{\;}{\sim}{\;}12r$ and $u{\;}{\sim}{\;}v$, for which $t_{c}{\;}{\sim}{\;}12t_{D}$ and $t_{R}{\;}{\sim}{\;}1728(M/m)^{2}t_{D}$. Internal irregularities associated with changes in the bulk potential might entail $t_{c}{\;}{\sim}{\;}0.4t_{D}$ and a large range of values for $t_{R}$.

FIGURE 15 exhibits the effects of Ornstein-Uhlenbeck noise with $t_{c}=4t_{D}$ and variable $t_{R}$ ranging from $10^{3}t_{D}$ and $10^{6}$. FIGURE 16 exhibits the effects of Ornstein-Uhlenbeck noise with $t_{c}=12t_{D}$ and $t_{R}=10^{3}t_{D}$ and $10^{4}t_{D}$. It is clear that, for $t_{c}=4t_{D}$, $t_{R}=10^{5}t_{D}$ has an appreciable and $10^{4}t_{D}$ a large effect. Similarly, for $t_{c}=12t_{D}$, a relaxation time $t_{R}=10^{4}t_{D}$ is short enough to be important and $10^{3}t_{D}$ again has a large effect. It thus follows that other giant galaxies with $m{\;}{\sim}{\;}M$ may be expected to have significant effects in a dense cluster environment where $d$ is as small as ${\sim}{\;}10r$, and that companion/satellite galaxies with $m{\;}{\sim}{\;}0.1M$ or even smaller could also play an important role.

FIGURE 17 exhibits the effects of Ornstein-Uhlenbeck noise with $t_{c}= 0.4t_{D}$ and variable $t_{R}$. Here it is clear that for $t_{R}$ as small as $10^{6}t_{D}$ the noise can have an appreciable effect. This would suggest that comparatively low level time-dependent irregularities associated with a galaxy out of equilibrium could indeed play a significant role in accelerating the approach towards a near-equilibrium.

What are the dynamical consequences of these external perturbations? Even though they are characterised by timescales which are long compared with $t_{D}$ they can have a substantial effect. For fixed amplitude, coloured noise with autocorrelation time $t_{c} \la t_D$ will have a much stronger effect than noise with $t_{c} \gg t_{D}$. However, external perturbations associated with nearby massive objects {\em will} in general have much larger amplitude and, at least in the outer portions of the galaxy, $t_{c}$ may not be that much longer than $t_{D}$. Perturbations with $t_{c}$ long compared with $t_{D}$ can still be important as an evolutionary mechanism if the amplitude is sufficiently large. This would suggest that, even if viable for isolated galaxies, `partially mixed' equilibria are {\em not} an option in high density environments!

\subsection{Strong perturbations (violent relaxation)}

The discussion hitherto has focused on orbit ensembles evolved in a fixed potential, possibly perturbed by low amplitude perturbations. However, chaotic phase mixing also has important potential implications for the evolution of systems that evidence a strong time dependence, such as galaxies in the process of merging. Indeed, one might argue (e.g., Kandrup, Vass, and Sideris 2003) that chaotic phase mixing is an important component which must be incorporated into any successful theory of violent relaxation. Arguably, the crucial ingredient of Lynden-Bell's (1967) original proposal is that phase mixing induces a coarse-grained approach towards equilibrium, his prototypical example being the motion of particles in a one-dimensional `pig-trough,' in which all the orbits are regular. The problem, however, with this and other similar examples is that the approach towards equilibrium is much less efficient than what simulations suggest for real stellar systems (cf. Kandrup 1999), typically proceeding at best as a power law in time. If, however, one considers a flow that is chaotic rather than regular, the approach towards equilibrium should proceed exponentially in time, which implies that a near-equilibrium can be achieved much more quickly.

One might perhaps object to this argument on the grounds that realistic systems will in general contain significant numbers of both regular and chaotic orbits, and that chaotic phase mixing should dominate the evolution of a time-dependent system only if chaotic orbits are much more common than regular orbits. Otherwise chaotic phase mixing would not be sufficiently ubiquitous as to have a dramatic effect on the system as a whole. It is certainly likely that, if only a small fraction of the orbits are chaotic, chaotic phase mixing will only have a comparatively minor effect. However, there is solid reason to believe that {\em time-dependent potentials tend to admit much larger numbers of chaotic orbits, especially if the potential exhibits pulsations.}

Theoretically it is easy to understand why this might be so. If an orbit is regular, it must be restricted by constants of the motion, either global integrals like angular momentum associated with global symmetries, or `local' integrals (cf. Lichtenberg \& Lieberman 1992) which restrict the motion of regular orbits in nonintegrable systems. Making the potential time-dependent removes the symmetry associated with time translation invariance so that neither the energy nor the Jacobi integral is conserved. In a generic time-independent potential, regularity requires two local integrals; in a generic time-dependent system three local integrals are required. One might thus anticipate that, at any given time, a larger fraction of the orbits will be chaotic, provided at least that the time-dependence is sufficiently strong that the energy cannot be treated as (nearly) an adiabatic invariant.

For the case in which the time-dependence involves large amplitude systematic oscillations, one can in fact demonstrate that, in many cases -- probably generically -- the relative measure of chaotic orbits and the size of the largest finite time Lyapunov exponent will both increase dramatically (Kandrup, Vass, and Sideris 2003). Indeed, it has been recognised in both the accelerator dynamics (cf. Gluckstern 1994) and the nonneutral plasma communities (cf. Strasburg and Davidson 2000) that chaos associated with such a resonance can be important at a practical level. In particular, it can result in particles in the core of an accelerator beam being ejected into an outerlying halo, thus resulting in a highly undesirable increase in the size of the beam. 

The basic idea is that the introduction of an oscillatory time-dependence can trigger a parametric resonance, involving a coupling between the frequencies associated with the pulsations and the frequencies of the orbits, making many/most of the orbits chaotic. The important point then is that, even for moderate amplitude oscillations, this resonance can be very broad, requiring only that the two sets of frequencies be comparable to within an order of magnitude or so. Given, however, that there is only one natural time scale for the problem, namely the dynamical $t_{D}$, which sets both the characteristic orbital time scale and the pulsation time, one would anticipate that, throughout most of the galaxy, the frequencies will be sufficiently close to trigger the resonance. Simple toy models involving an integrable Plummer sphere subjected to damped oscillations can, within a time ${\sim}{\;}10t_{D}$, exhibit both near-complete chaotic phase mixing and an approach towards a nearly time-independent state (Kandrup, Vass, and Sideris 2003).

The same physical processes can also act on shorter scales. A supermassive black hole binary introduced into the center of a galaxy -- with or without a cusp -- serves as a time-dependent perturbation which can trigger large amounts of chaos even at distances from the center much larger than the binary orbit; and the resulting chaos can lead to efficient phase mixing which occasions significant changes in the density distribution and, hence, the observable surface brightness profile (Kandrup, Sideris, Terzi{\'c}, and Bohn 2003).
\vskip .2in
The work described here and in Paper I reinforces the expectation that chaos may be ubiquituous in cuspy galaxies, especially those that manifest significant deviations from axisymmetry, and that that chaos may have significant implications for both structure and evolution. In particular, `realistic' galaxies, characterised by a bulk potential that admit a complex coexistence of regular and chaotic behaviour, may be substantially more susceptible to various `irregularities' than had been recognised originally. The crucial remaining issue, which has yet to be addressed, would seem to be precisely how the effects described here will manifested in the context of a time-dependent, fully self-consistent evolution. 

Finally, however, one can conclude by comparing the results derived in
this paper with models which might naively seem to contradict one of its
principal conclusions (as well as of Gerhard \& Binney [1985], Merritt \&
Fridman [1996], Merritt \& Valluri [1996], and Merritt [1997]), namely
that chaos should be ubiquituous in cuspy triaxial galaxies. Specifically,
Holley-Bockelmann {\em et al} (2001) (hereafter HB) constructed $N$-body
realisations of triaxial (quasi-)equilibria with a central density cusp by
adiabatically `squeezing' spherical ${\gamma}=1$ Hernquist models and, for
one particular choice of moderately triaxial axis ratios ($a:b:c =
1:0.85:0.7$) found no chaotic, or at least no strongly chaotic, orbits.
And similarly, Poon \& Merritt (2002) (hereafter PM) were able to construct
Schwarzschild models with $\gamma=1$ and 2 profiles containing only regular
orbits (although they also constructed models with both regular and chaotic
orbits). $N$-body realisations of all their Schwarzschild models revealed
only small fluctuations over a few crossing times.

In point of fact it is not completely clear that there are no chaotic orbits in
the HB model. As those authors themselves stated, their
algorithm to identify chaotic orbits, based on their Fourier transforms,
could easily have missed a large number of nearly-regular, but still
chaotic, orbits. Moreover, since their search for chaos involved
integrating initial conditions in the fixed $N$-body background associated
with a single $t=$ constant snapshot, they excluded explicitly the
possibility that weak amplitude oscillations which, as is evident from
Figure 3 of HB, exist in the real model, could trigger chaos via
parametric resonance. This model is, however, clearly interesting in that
it may represent an example of a near-equilibrium supported using only
`local' (i.e., `third') integrals. All three of these points can also be
made for the PM models.

In any event, it could well be that neither the triaxial Dehnen models nor
the HB or PM models constitute true time-independent self-consistent
equilibria. However, as emphasised in Paper I (and discussed more
extensively in Kandrup [2002]), it is quite possible that a galaxy will
settle initially into one triaxial near-equilibrium state and then evolve
slowly through a sequence of near-equilibria without necessarily becoming
more axisymmetric. There is at the present time no compelling evidence
that a slow secular evolution {\em necessarily} involves an evolution
towards axisymmetry. What might, however, be true is that a galaxy that
resembles a triaxial Dehnen model with large measures of wildly chaotic
orbits would evolve towards less chaotic configurations better represented
by an HB- or PM-type model. The work described in the present paper could
provide potentially significant insights as to precisely how such an
evolution might proceed.

\section*{Acknowledgments}

HEK was supported in part by NSF AST-0070809 at the University of Florida.
CS was supported by NASA NAG 5-8238 at the University of Michigan. CS
would like to thank Kelly Holley-Bockelmann for her help with answering
questions regarding Holley-Bockelmann et al (2001). HEK would like to
thank his colleagues who, wittingly or not, provided much of the computer
time required for the approximately 800,000 orbits computed in the course
of the preparation of this paper.

\vfill\eject
\end{document}